\documentclass[runningheads]{llncs}
\usepackage{graphicx}
\usepackage[T1]{fontenc}
\usepackage[pagebackref=true]{hyperref}
\usepackage{orcidlink}
\newcommand {\dpol} {\mathcal{DP} }

\newcommand {\p} {\mathcal{P} }
\newcommand {\s} {\mathcal{S} }
\newcommand {\ct} {\mathcal{T} }
\newcommand {\f} {\mathcal{F} }
\newcommand {\fq} {\mathbb{F}_q }
\newcommand {\fqn} {\mathbb{F}_q^n }
\newcommand {\fqm} {\mathbb{F}_q^m }
\usepackage{amsmath}
\usepackage{amssymb,mathtools}
\usepackage{longtable}
\usepackage{setspace}
\usepackage{algorithm}%
\usepackage{algorithmicx}%
\usepackage{algpseudocode,booktabs}%

\usepackage[paragraphs, wordspacing, tracking, floats, mathspacing, subtle]{savetrees}

\usepackage{adjustbox}

\begin{document}

\newif\ifsubmission
%\submissiontrue
\submissionfalse

\title{VDOO: A Short, Fast, Post-Quantum Multivariate Digital Signature Scheme}
\titlerunning{VDOO signature scheme}
% If the paper title is too long for the running head, you can set
% an abbreviated paper title here
%
\ifsubmission

\else
\author{Anindya Ganguly \and Angshuman	Karmakar\orcidlink{0000-0003-2594-588X} \and Nitin Saxena \orcidlink{0000-0001-6931-898X}} %\and
%Third Author\inst{3}\orcidID{2222--3333-4444-5555}}
%%
%\authorrunning{F. Author et al.}
%% First names are abbreviated in the running head.
%% If there are more than two authors, 'et al.' is used.
\institute{Department of Computer Science and Engineering, IIT Kanpur \\\email{\{anindyag,angshuman,nitin\}@cse.iitk.ac.in}}
\fi
% \author{First Author\inst{1}\orcidID{0000-1111-2222-3333} \and
% Second Author\inst{2,3}\orcidID{1111-2222-3333-4444} \and
% Third Author\inst{3}\orcidID{2222--3333-4444-5555}}
% %
% \authorrunning{F. Author et al.}
% % First names are abbreviated in the running head.
% % If there are more than two authors, 'et al.' is used.
% %
% \institute{Princeton University, Princeton NJ 08544, USA \and
% Springer Heidelberg, Tiergartenstr. 17, 69121 Heidelberg, Germany
% \email{lncs@springer.com}\\
% \url{http://www.springer.com/gp/computer-science/lncs} \and
% ABC Institute, Rupert-Karls-University Heidelberg, Heidelberg, Germany\\
% \email{\{abc,lncs\}@uni-heidelberg.de}}
% %
\maketitle              % typeset the header of the contribution
\begin{abstract}
% A sizeable quantum computer combined with Shor's polynomial time integer factorization algorithm will render most of our public-key infrastructure based on integer factorization or Elliptic-curve discrete logarithm problem unsafe. Hence, we need to replace existing cryptosystems with quantum-secure of post-quantum cryptosystems. 

Hard lattice problems are predominant in constructing post-quantum cryptosystems. However, we need to continue developing post-quantum cryptosystems based on other quantum hard problems to prevent a complete collapse of post-quantum cryptography due to a sudden breakthrough in solving hard lattice problems. Solving large multivariate quadratic systems is one such quantum hard problem. %Compared to lattice-based cryptography, cryptosystems based on multivariate systems have considerably smaller key sizes. 

Unbalanced Oil-Vinegar is a signature scheme based on the hardness of solving multivariate equations. In this work, we present a post-quantum digital signature algorithm VDOO (Vinegar-Diagonal-Oil-Oil) based on solving multivariate equations. We introduce a new layer called the diagonal layer over the oil-vinegar-based signature scheme Rainbow. This layer helps to improve the security of our scheme without increasing the parameters considerably. Due to this modification, the complexity of the main computational bottleneck of multivariate quadratic systems i.e. the Gaussian elimination reduces significantly. Thus making our scheme one of the fastest multivariate quadratic signature schemes. Further, we show that our carefully chosen parameters can resist all existing state-of-the-art attacks. The signature sizes of our scheme for the National Institute of Standards and Technology's security level of I, III,  and V are 96, 226, and 316 bytes, respectively. This is the smallest signature size among all known post-quantum signature schemes of similar security.
\end{abstract}
%
%
%
%\vspace{-15pt}
\section{Introduction}\label{sec:intro}
%\vspace{-8pt}
Cryptography is the study of different methods to safeguard our sensitive information in the ever-expanding digital world. The security assurances of cryptographic schemes especially public-key cryptographic schemes emanate from the computational intractability of some underlying hard problems. Currently, public-key cryptographic schemes such as Rivest-Shamir-Adleman~\cite{rivest1978method}, elliptic-curve discrete logarithm~\cite{miller1985use} are predominant in our public-key infrastructure. However, in the context of the rapid development of quantum computers, these schemes exhibit a significant drawback. The underlying hard problems of these schemes \textit{i.e.} integer factorization and discrete logarithm problem can be solved \emph{easily} due to the polynomial time quantum algorithms developed by Shor~\cite{shor1994algorithms} and Proos-Zalka~\cite{Proos_Zalka_2003} respectively. Therefore, quantum-resistant hard problems have gained popularity among designers for designing public-key cryptosystems for the future. A landmark event in the development of such quantum-resistant or post-quantum cryptography (PQC) is the PQC standardization procedure~\cite{chen2017nist} initiated by the National Institute of Standards and Technology (NIST) to select quantum-safe cryptographic primitives such as key encapsulation mechanisms (KEM), public-key encryption (PKE), and digital signature algorithm.
%The mathematical art of cryptography safeguards the swiftly expanding digital world. Presently, RSA \cite{rivest1978method}, ECDSA and ECDH \cite{miller1985use} are used extensively throughout the globe. However, these schemes have a significant drawback. They lack quantum security. The quantum algorithm developed by Shor poses a threat to these classical cryptosystems \cite{shor1994algorithms}. Quantum-safe assumptions are of interest to designers considering this issue. NIST called for the standardization of quantum-safe public key encapsulation mechanisms (KEM) and digital signature algorithms in 2016 \cite{chen2017nist}.
In 2022, NIST standardized~\cite{nist_final_report} one KEM (Crystals-Kyber~\cite{Kyber-Kem}) and three signature schemes (SPHINCS+~\cite{web:sphincs}, Crystals-Dilithium~\cite{dilithium}, and Falcon~\cite{web:falcon}) after rigorous scrutiny spanning multiple years. Among these only SPHINCS+ is based on the hardness of cryptographically secure hash functions, while Crystals-Kyber (KEM), Crystals-Dilithium, and Falcon are based on hard lattice problems. As the majority of these constructions are lattice-based, there is a lingering risk that a breakthrough in the cryptanalysis of lattice-based cryptography can reduce the security of these schemes drastically. Thus putting the whole plan to migrate to post-quantum cryptography in jeopardy. Such incidents are not uncommon. Recently, Decru et al.~\cite{decru_sidh} proposed an attack to completely break the security of supersingular isogeny Diffie-Hellman~\cite{sidh} which was earlier considered quantum-safe and was also a finalist in the NIST's standardization procedure. Therefore, it is prudent to diversify the portfolio of different quantum-safe problems for seamless migration to a post-quantum world. There exist other problems that are considered quantum-safe, such as multivariate quadratic (MQ)~\cite{patarin1997oil,kipnis1999unbalanced}, isogeny-based~\cite{de2020sqisign}, and code-based~\cite{bernstein2017classic}. Standardizing cryptographic primitives necessitates a rigorous and comprehensive investigation. NIST reissued a call~\cite{nistadditional} for quantum-safe signature schemes to standardize some more signature schemes to diversify the portfolio of quantum-resistant schemes. Due to its small signature size, multivariate oil-vinegar construction has gained significant attention during this standardization process.

Multivariate cryptography relies on the intractability of root findings of MQ equations. The goal of the MQ problem is to find a solution to a system of multivariate quadratic polynomials in the finite field $\fq$. In other words, the hardness classification of this problem is NP-hard~\cite{johnson1979computers}. Numerous schemes, such as Matsumoto-Imai encryption scheme~\cite{matsumoto1988public}, Oil-Vinegar~\cite{patarin1997oil} signature, Rainbow~\cite{ding2005rainbow} signature, Triangular~\cite{moh1999public,shamir1994efficient,yang2005building} signature, Simple Matrix encryption \cite{tao2013simple}, and Mayo \cite{beullens2022mayo}, have been developed based on multivariate cryptography. Patarin first proposed the Oil-Vinegar signature~\cite{patarin1997oil}. A successful forgery attack was shown by Kipnis and Shamir~\cite{kipnis1998cryptanalysis} against this scheme. Further, Kipnis, Patarin, and Goubin upgraded the signature scheme by proposing Unbalanced Oil-Vinegar (UOV)~\cite{kipnis1999unbalanced}. 

%Ding and Schmidt proposed Rainbow in 2005~\cite{ding2005rainbow}. 
Rainbow was a third-round NIST candidate~\cite{ding2005rainbow}, which is the first multi-layer construction based on unbalanced oil-vinegar.
%One of the major contribution The main interesting part of Rainbow~\cite{ding2005rainbow} by Ding et al. is the multi-layer approach. The ground layer is the grand old scheme of unbalanced oil-vinegar~\cite{kipnis1999unbalanced}. Rainbow is comparatively more efficient than UOV due to its construction. This scheme was submitted to NIST-competition~\cite{chen2017nist} and appeared as a third-round candidateDue to the inclusion in the NIST competit
Therefore, the cryptanalysis of Rainbow has been a well-studied area for the last decade. This resulted in many new novel attacks such as direct attack~\cite{bardet2002algebraic,faugere1999new,faugere2002new}, min-rank attack~\cite{billet2006cryptanalysis,bardet2002algebraic,bardet2020improvements,baena2022improving}, band-separation attack~\cite{ding2008new,thomae2012generalization,smith2020rainbow}, rectangular min-rank and intersection attack~\cite{beullens2021improved}. In 2023, Beullens proposed a cryptanalysis and reduced the security of Rainbow significantly. Rainbow team suggested using the old SL-3 (high security) parameter set as new SL-1 (low security) parameters~\cite{nistgoogle} to mitigate the attack. As Beullens' attack only applies to the Rainbow structure, therefore building scheme on the top of the oil-vinegar layer is still believed to be secure.

In 2022, Cartor \textit{et al.} internally perturbed the second layer of Rainbow by mixing oil variables quadratically \cite{cartor2022iprainbow}. However, this mixing significantly increased the signature generation time. Also, parameter sets proposed by designers are not practical in terms of efficiency. Therefore, designing a new signature scheme that can resist the simple attack while being practical, is an interesting open problem. %novel research direction.

\subsection{Our Contribution and Motivation} 
%\vspace{-5pt}
In the context of this endeavor, we summarize our contributions below. 
\begin{itemize}
    \item We review related multivariate signature schemes and provide a comprehensive analysis of their design and performance in Section~\ref{sec:background}.
    \item We present Vinegar-Diagonal-Oil-Oil (VDOO), a novel multivariate signature scheme based on unbalanced oil and vinegar in Section~\ref{sec:scheme}. Compared to other UOV schemes VDOO boasts three primary benefits: {\it simplicity}, {\it efficiency}, and {\it security} (see Sections~\ref{sec:sec_analysis_params} and~\ref{sec:parms}). To the best of our knowledge, we are the first to introduce a diagonal layer within the UOV framework, demonstrating that it enhances efficiency without compromising security. 
    \item We establish that VDOO effectively withstands all current attacks and outline the EUF-CMA security of our scheme. Through meticulous parameter selection, our findings reveal that it achieves a remarkably compact smallest signature size of 96 bytes (see Sections~\ref{sec:sec_analysis_params} and~\ref{sec:parms}), contrasting favorably with NIST-standardized post-quantum signatures (Crystals-dilithium \cite{dilithium}, Falcon \cite{web:falcon}, and SPHINCS+ \cite{web:sphincs}).
\end{itemize}
%\subsubsection{Conceptually simple: new design components.} 
\subsubsection{Introduction of a new simple design element.} 
\begin{figure}[!ht]
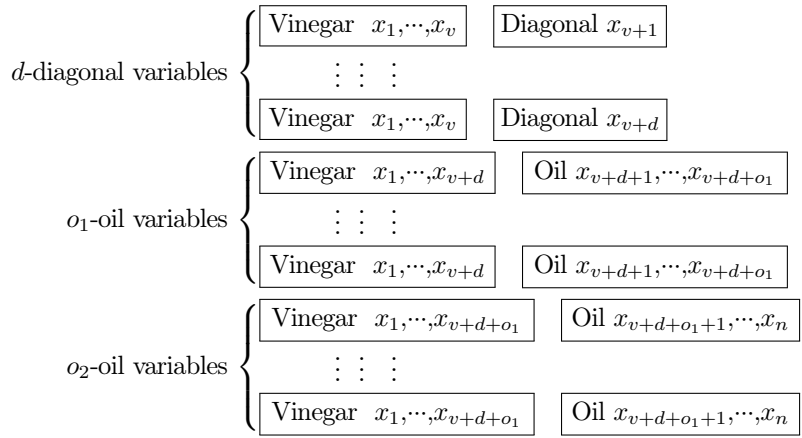

\centering
 \begin{align*}
       d\mbox{-diagonal variables }&
    \begin{cases}
       \framebox[1.1\width]{Vinegar~~$x_1,\cdots,x_v$}& \framebox[1.1\width]{Diagonal $x_{v+1}$}\\ \hspace{1cm} \vdots ~~\vdots ~~~\vdots \\
      \framebox[1.1\width]{Vinegar~~$x_1,\cdots,x_v$}& \framebox[1.1\width]{Diagonal $x_{v+d}$} 
    \end{cases}\\ 
       o_1\mbox{-oil variables }&
    \begin{cases}
       \framebox[1.1\width]{Vinegar~~$x_1,\cdots,x_{v+d}$}& \framebox[1.1\width]{Oil $x_{v+d+1},\cdots,x_{v+d+o_1}$}\\\hspace{1cm} \vdots ~~\vdots ~~~\vdots \\
     \framebox[1.1\width]{Vinegar~~$x_1,\cdots,x_{v+d}$}& \framebox[1.1\width]{Oil $x_{v+d+1},\cdots,x_{v+d+o_1}$}
    \end{cases} \\
      o_2 \mbox{-oil variables }&
     \begin{cases}
       \framebox[1.1\width]{Vinegar~~$x_1,\cdots,x_{v+d+o_1}$}& \framebox[1.1\width]{Oil $x_{v+d+o_1+1},\cdots,x_{n}$}\\ \hspace{1cm} \vdots ~~\vdots ~~~\vdots \\
     \framebox[1.1\width]{Vinegar~~$x_1,\cdots,x_{v+d+o_1}$}& \framebox[1.1\width]{Oil $x_{v+d+o_1+1},\cdots,x_{n}$}
    \end{cases}
  \end{align*}\caption{Variables in each layer of the VDOO central map}\label{fig: vdoocent}
\end{figure}

VDOO is a new layer-based construction, which has one diagonal layer and then two UOV layers. We are adding each new variable in the central polynomial one by one diagonally. This offers efficiency. 
%Our trick reduces the cost of Gaussian elimination, which is the primary computational bottleneck resides in  multivariate signature scheme .
%Our tricks reduces the cost of the Gaussian elimination (shortly $\mathsf{GE}_{(q,n)}$ 
This translates to a reduction of the Gaussian elimination ($\mathsf{GE}_{(q,n)}$
\footnote{$\mathsf{GE}_{(q,n)}$: Gaussian elimination on a linear system with $n$ unknowns and $n$ linear equation over $\fq$.}) which is the major computational bottleneck in the signature generation process.
%This trick reduces the size of the linear system when Gaussian elimination is applied. Earlier, three-layer Rainbow needs three Gaussian elimination, however, our proposal needs two Gaussian elimination.The following structure illustrates the construction.  
Suppose $x_1,x_2,\cdots,x_v,x_{v+1},\cdots,x_{v+d},\cdots,x_{v+d+o_1},\cdots,$ $x_{v+d+o_1+o_2=:n}$ are $n$ variables defined over $\fq$. In our construction, we call first $v$-variables as {\it vinegar variables}, next $d$-variables as {\it diagonal variables}, then next $o_1$ variables are {\it first-layer oil variables}, and last $o_2$ variables are {\it second-layer oil variables}. Figure \ref{fig: vdoocent} illustrates the distribution of the variables in each layer of the VDOO central polynomial map.
%\begin{adjustbox}{width=\columnwidth,center}\adjustbox{max width=\textwidth}{%
%\end{adjustbox}

% \iffalse
% \smallskip\noindent\textbf{Efficiency. }
% To thwart Beullens' simple attack~\cite{beullens2022breaking}, Rainbow augmented the parameter set, which results in inefficiencies. The increase in Rainbow's parameters subsequently escalates the Gaussian elimination cost, as its complexity becomes approximately $o_1^3 + o_2^3$ where $o_1$ and $o_2$ are parameters of Rainbow \cite{ding2005rainbow}. 
% In our scheme, we adapt $d \approx (o_1 + o_2)/3$, $o_1' \approx (o_1 + o_2)/3$, and $o_2' \approx (o_1 + o_2)/3$ as the new parameters. This adjustment results in a Gaussian elimination complexity of around ${o_1}'^3 + {o_2}'^3$. To illustrate, consider the signature generation process for SL-1 parameters: UOV necessitates $\mathsf{GE}_{(256,64)}$, Rainbow requires $\mathsf{GE}_{(256,32)}$ and $\mathsf{GE}_{(256,48)}$, while VDOO needs only $\mathsf{GE}_{(16,34)}$ and $\mathsf{GE}_{(16,36)}$ (for further details, refer to Table \ref{tab:compmul}). Consequently, this modification notably impacts the scheme's performance.
% \fi

\smallskip\noindent\textbf{Efficiency. }
To thwart Beullens' simple attack~\cite{beullens2022breaking}, the authors of Rainbow increased the parameter set \cite{nistgoogle}, which results in increasing the Gaussian elimination cost. The complexity of Gaussian elimination becomes approximately $o_1^3 + o_2^3$ where $o_1$ and $o_2$ are number of oil variables of Rainbow~\cite{ding2005rainbow}. 
In our scheme, we adapt $d \approx (o_1 + o_2)/3$, $o_1' \approx (o_1 + o_2)/3$, and $o_2' \approx (o_1 + o_2)/3$ as the new parameters. This adjustment results in a Gaussian elimination complexity of around ${o_1}'^3 + {o_2}'^3$. To illustrate, consider the signature generation process for security level one (SL-1) parameters~\cite{chen2017nist}: UOV requires $\mathsf{GE}_{(256,64)}$, Rainbow requires $\mathsf{GE}_{(256,32)}$ and $\mathsf{GE}_{(256,48)}$, while VDOO needs only $\mathsf{GE}_{(16,34)}$ and $\mathsf{GE}_{(16,36)}$ (for further details, refer to Table \ref{tab:compmul}). Consequently, this modification notably improves our scheme's performance.

%To prevent the Beullens simple attack, Rainbow team suggested increasing the parameter set to resist the attack, which causes inefficiency. Main computational bottleneck for multivariate schemes is Gaussian elimination, $\mathsf{GE}_{(q,n)}$ \footnote{$\mathsf{GE}_{(q,n)}$: Gaussian elimination on a linear system with $n$ unknowns and $n$ linear equation over $\fq$. This computation needs $O(n^3)$-field operations.}. Incresing parameters of Rainbow increased cost of the Gaussian elimination. Since the complexity of Gaussian elimination is $\approx N^3$, therefore for Rainbow it is $o_1^3+o_2^3$. Now for our scheme, the $d\approx (o_1+o_2)/3$, $o_1'=(o_1+o_2)/3$, and $o_2'=(o_1+o_2)/3$. Therefore, the complexity for the Gaussian elimination is approximately $({o_1}'^3+{o_2}'^3)$. For example, to design SL-1 parameters, UOV needs to perform $\mathsf{GE}_{(256,64)}$, Rainbow needs $\mathsf{GE}_{(256,32)}$, $\mathsf{GE}_{(256,48)}$. However, VDOO requires $\mathsf{GE}_{(16,34)}$ and $\mathsf{GE}_{(16,36)}$ (more details see \ref{tab:compmul}). Thus this modification affects the performance significantly.
%\vspace{-5pt}
\smallskip\noindent\textbf{Resistance to existing attacks.} 
%We analyze that all strong attacks (available in multivariate cryptography) can apply to our scheme also.To retrieve diagonal variables attacker tries to remove top oil layers first. Beullens' proposal helps attacker to remove these layers. To break our round-one parameter set, the simple attack requires $2^{134}$-field operations. Beullens also combined the simple attack with the rectangular min-rank attack. Like earlier, we perform the combined attack against our scheme. We find that the attack needs $2^{138}$ field operations for our level one parameter set. We also perform intersection attack, direct attack on our scheme and found attack the complexity is above $2^{134}$-field operations. Therefore, these references suggest that VDOO is seems to be secure against all known attacks. We also find VDOO is EUF-CMA security of our scheme.
We comprehensively analyze all possible attacks on multivariate cryptographic schemes against our scheme. In an attempt to recover diagonal variables, potential attackers begin by eliminating the uppermost oil layers. Beullens proposed method \cite{beullens2022breaking} facilitates the removal of these layers, aiding attackers. For instance, in order to compromise our round-one parameter set, a straightforward attack necessitates $2^{134}$-field operations. Furthermore, Beullens combined this simple attack with the rectangular min-rank attack~\cite{beullens2021improved,beullens2022breaking}. In line with previous efforts, we execute this combined attack against our scheme, determining that it requires $2^{138}$-field operations to break SL-1 parameter set. Additionally, we conduct the intersection attack and the direct attack on our scheme, both of which exhibit complexities exceeding $2^{134}$-field operations. Consequently, these references collectively imply that VDOO appears to withstand all known attacks securely. We also outline the EUF-CMA security of the VDOO scheme.

%\vspace{-4pt}
\smallskip\noindent\textbf{Small signature size. } %The aforementioned attack assists us in formulating parameter tuples for various security levels. 
We present multiple parameters that can withstand the aforementioned attacks. Specifically, our level-one parameters that can provide $128$-bit classical and $96$-bit post-quantum security has a signature size of 96 bytes and public-key size of 238KB (further elaborated in Table \ref{tab:vdoo}). %Specifically, our round one parameter set, designed to achieve a minimum of 128-bit security, necessitates a public key of 238KB in size (further elaborated in Table \ref{tab:vdoo}). 
This is the smallest signature size among the majority of all multivariate signature schemes (for additional insights, refer to Tables \ref{tab:compmul} and \ref{tab:compsign}).

\ifsubmission
\else
\smallskip{\textbf{\textit{Roadmap.}}} 
In the upcoming Section \ref{sec:background} we present a generic construction of multivariate signatures, and some earlier results. Section \ref{sec:scheme} proposes a new post-quantum multivariate signature scheme called VDOO. The cryptanalysis of our scheme is presented in Section \ref{sec:sec_analysis_params}. In Section \ref{sec:parms}, we give the parameters for different security levels and we also compare our results with the state-of-the-art. Our section \ref{sec:conclusion} presents conclusions and explores potential future directions for our work.
\fi
%\vspace{-10pt}
\section{Prior Results}\label{sec:background}
In this section, we introduce some essential mathematical notations and symbols. We then provide a generic construction for multivariate signatures. Following that, we outline the central polynomial for UOV and Rainbow \cite{kipnis1999unbalanced,beullensuov,ding2005rainbow}. Additionally, we describe the subspace representation of Rainbow \cite{beullens2021improved}, which is particularly valuable for cryptanalysis purposes. Next, we cover recent multivariate signature schemes \cite{beullens2022mayo,furue2023qr,faugereprov,ding2023tuov,voxfranceprincipal} that were submitted as part of the NIST additional round for post-quantum signature standardization \cite{nistadditional}. Finally, we present the required hardness assumptions for these multivariate signatures to understand their cryptanalysis.

\iffalse
\subsection{Notations} Let, $\fq$ be the finite field with $q$ elements.  We define two affine maps $\mathcal{S}:\mathbb{F}_q^m\rightarrow \mathbb{F}_q^m$ and $\mathcal{T}:\mathbb{F}_q^n\rightarrow \mathbb{F}_q^n$, and one quadratic map $\mathcal{F}=(f_1,\cdots,\ f_m):\mathbb{F}_q^n\rightarrow \mathbb{F}_q^m$. We denote $[n]$ for the set $\{1,\ 2,\ \cdots,\ n\}$ and $[i:j]$ denotes $\{i,\ i+1,\cdots,\ j\}$. We use `bold' to identify vectors, e.g. $\mathbf{x}$ is a vector and $x$ is a field element. 
\fi
%\subsubsection*{Notations:}
\noindent
\textbf{Notations. }Let, $\fq$ be the finite field with $q$ elements.  We define two invertible affine maps $\mathcal{S}:\mathbb{F}_q^m\rightarrow \mathbb{F}_q^m$ and $\mathcal{T}:\mathbb{F}_q^n\rightarrow \mathbb{F}_q^n$, and one quadratic map $\mathcal{F}=(f_1,\cdots,f_m):\mathbb{F}_q^n\rightarrow \mathbb{F}_q^m$. We denote $[n]$ for the set $\{1,2,\cdots,n\}$ and $[i:j]$ denotes $\{i,i+1,\cdots,j\}$. We use lowercase and bold lowercase alphabets to denote field elements and vectors respectively. The notation $a\in_U S$ is used to interpret \emph{$a$ is a random element in the set $S$}.
%We use `bold' to identify vectors, e.g. $\mathbf{x}$ is a vector and $x$ is a field element. 
%\vspace{-10pt}
%\subsection{General Multivariate Signature Schemes}
%A multivariate signature scheme relies on a quadratic homogeneous system as its foundation. 
\subsection{Generic Multivariate Signature Schemes}
Here we briefly describe a generic construction for multivariate signature schemes. Due to the NP-hardness of inverting a randomly generated quadratic system~\cite{johnson1979computers}. However, signers can leverage a specially structured quadratic system to efficiently perform the inversion. This specialized system is commonly referred to as the \emph{central map} and is typically denoted as $\mathcal{F} = (f_1, \cdots, f_m)$, where each $f_i$ represents a specifically structured multivariate quadratic polynomial. Signers must conceal this unique structure from third parties to prevent forgery attacks. To achieve this objective, signers employ one or two random invertible linear maps: $\s$ and $\ct$.
%$\mathcal{S}: \mathbb{F}_{q}^{m} \longrightarrow  \mathbb{F}_{q}^{m}$ and $\ct : \mathbb{F}_{q}^{n} \longrightarrow \fqn$ 
%(where each coefficient is randomly chosen). 
Consequently, the public key is constructed by composing these linear maps along with the central map, denoted as $\p=~\s\circ\f\circ \ct~:~\fqn\longrightarrow~ \fqm$.

The secret key comprises $ \s, ~\ct$ and $\mathcal{F}$. A hash function, denoted as $\mathcal{H}: \{0,1\}^* \longrightarrow \mathbb{F}_q^m$, is employed to generate a vector $\mathbf{m} \in \fqm$ from a message $msg \in \{0,1\}^*$. The signature generation process unfolds as follows: first, compute $\mathbf{d} \leftarrow \mathcal{S}^{-1}(\mathbf{m})$, then $\mathbf{d}' \leftarrow \mathcal{F}^{-1}(\mathbf{d})$, and finally $\mathbf{s} \leftarrow \ct^{-1}(\mathbf{d}')$. The signer sends the signature $\mathbf{s}$ for the message $msg$ to the verifier. The verifier simply evaluates the polynomial map $\p$ on $\mathbf{s}$ and checks whether it matches the hash of the message, i.e., whether $\mathbf{m} = \p(\mathbf{s})$ holds or not.
%\vspace{-10pt}
\subsection{Unbalanced Oil-Vinegar (UOV)}
%The Oil-Vinegar (OV) signature scheme was initially introduced by Patarin in 1997 \cite{patarin1997oil}. However, in 1998, Kipnis and Shamir discovered a vulnerability known as `invariant subspace attack' against this scheme \cite{kipnis1998cryptanalysis}. In response to this attack, Kipnis, Patarin, and Goubin made significant improvements by increasing the number of vinegar variables. This proposal is named as `unbalanced oil-vinegar signature scheme' \cite{kipnis1999unbalanced}. 

The Oil-Vinegar (OV) signature scheme was initially introduced by Patarin~\cite{patarin1997oil}. However, due to the Kipnis-Shamir's~\cite{kipnis1998cryptanalysis} \emph{invariant subspace} attack, this scheme was modified by increasing the number of vinegar variables. This is known as the Unbalanced Oil-Vinegar (UOV) signature scheme~\cite{kipnis1999unbalanced}. 

Consider the OV central map, denoted as $\f$. Split all variables of $\mathbf{x}=(x_1,\cdots,x_v$ $,\cdots,x_n)$ into two buckets: the first bucket has first $v$ variables representing vinegar, and the second bucket contains next $o$ variables representing oil, where $n = v + o$ and $o=m$. To create a multivariate quadratic homogeneous polynomial, combine variables involving vinegar $\times$ vinegar and vinegar $\times$ oil, while excluding all oil $\times$ oil terms. 
\begin{definition}[OV Central Polynomial Map]
A central map $\f=(f_1,\cdots , f_m):\fqn\to\fqm$ is known as {\it OV central polynomial map} when each $f_i$ is of the form $f_i\;(\mathbf{x})~=~\sum_{i=1}^v\sum_{j=1}^n ~\alpha_{i,j}^{(k)}~x_i\;x_j$ where  $i\leq j,~k\in [v+1:n]$, $\mathbf{x}\in \fqn$, and $\alpha_{i,j}^{(k)}\in_U \fq$.
%$$f^{(k)}\;(\;x_1,x_2,\cdots,x_n\;)~=~\sum_{i=1}^v\sum_{j=1}^n ~\alpha_{i,j}^{(k)}~x_i\;x_j$$ where $i\leq j,~k\in [v+1:n]$ and $\alpha_{i,j}^{(k)}\in_U \fq$. 
\end{definition}\label{defn:OV}
Notably, if anyone randomly fixes vinegar variables, then the remaining part would be linear in the oil variables. Therefore, the quadratic system reduces to a linear system of $o$ linear equations with $o$ unknowns. 
%\vspace{-10pt}
\subsection{Rainbow} Rainbow is a multi-layer variant of UOV \cite{ding2005rainbow}. For simplicity consider a two-layer Rainbow. Suppose $n=v+o_1+o_2$, where the first $v$ variables are vinegar and the next $o_1$ and $o_2$ variables are the first and second layer of oil variables respectively. This can be viewed as a UOV map with $v+o_1$ variables and $o_1$ oil variables and the next layer $v+o_1+o_2$ variables and $o_2$ oil variables.
%In first layer, constructs a UOV map with $v+o_1$ variables and $o_1$ constraints and next layer constructs $v+o_1+o_2$ variables and $o_2$ constraints. 

%Among $n$ variables choose first $v$ variables as vinegar and  next $o_1$ oil variables to construct $o_1$ homogeneous quadratic equations. So when one fixes values of vinegar variables then the quadratic system reduces to a linear system with $ o_1$ variables and $o_1$ many constraints. Hence the Gaussian elimination can help to solve this linear system. Now in the second layer, all known $v+o_1$ variables are treated as vinegar variables and newly added $o_2$-many variables are oil variables. These variables help to construct $o_2$ many homogeneous quadratic equations. 

\begin{definition}[Rainbow Central Polynomial Map] The mathematical expression for 
$l$-layer Rainbow central polynomial is as follows. 
  $$f_k(x_1,x_2,\cdots,x_n)=\sum_{i,j\in [r];~ i\leq j} \alpha_{ij}^{(k)}x_ix_j+ \sum_{i\in [r]; ~j\in [r+1:r+o_r]}\beta_{ij}^{(k)}x_ix_j$$
   where for each $k\in [r+1:r+o_r]$, elements $\alpha_{ij}^{(k)}$,and $\beta_{ij}^{(k)}$ are taken from $\mathbb{F}_q $; $r$ denotes the layer, and $r\leq l$ where $l$ is the total number of layers in Rainbow.
   \end{definition}
\subsection{Beullens Subspace Description}
For a better view of cryptanalysis on Rainbow, Beullens explained the construction of Rainbow via subspaces~\cite{beullens2021improved}. Using this description, he derived the simple attack~\cite{beullens2022breaking}.  To elaborate this idea, initially, we define a differential polar form of a polynomial map.

The {\em differential polar map} of a polynomial map $\p$ is denoted by $\dpol:~\fqn\times\fqn~\rightarrow~\fqm$ and defined as $\dpol(\mathbf{x},~ \mathbf{w})=~\p(\textbf{x}+\textbf{w})-~\p(\textbf{x})-~\p(\textbf{w}).$ Note that, we only consider homogeneous quadratic polynomials, so throughout this paper, $\mathcal{P}(0)=0$.
\begin{figure}[h]
\centering
\includegraphics[scale=.6]{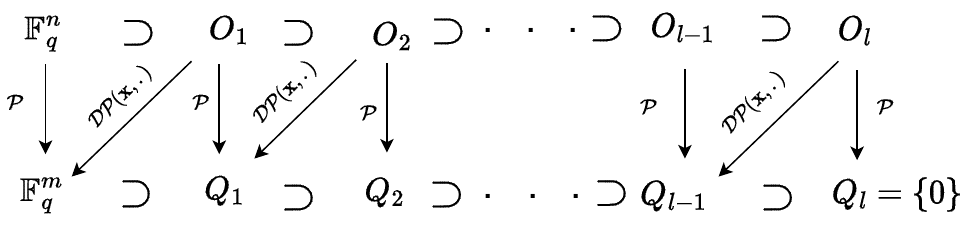}
\caption{$l$ layer Rainbow}    \label{fig: rainbow}
\end{figure}

\textbf{Trapdoor information.} This part describes the trapdoor information of $l$-layer Rainbow. At first, signer chooses a secret chain of nested subspaces: input subspaces $O_1\ \supset O_2\ \supset\cdots\ \supset O_l$ and output subspaces $Q_1\ \supset Q_2\ \supset\cdots\ \supset Q_l=\{0\}$. Using this secret, one can construct a public polynomial map as follows.
 \begin{itemize}
     \item  $\mathcal{P}$ maps each $O_i$ to $Q_i$ and\item  for any $\mathbf{x}\in_U\mathbb{F}_q^n$, $\mathcal{DP}_{\mathbf{x}}\colon O_i\to Q_{i-1} \,$ is a linear map (see Figure~\ref{fig: rainbow})
%$$\mbox{ for all } ~\mathbf{x}~\in ~\mathbb{F}_q^n \mbox{ and all } \mathbf{w}~\in ~O_i \;,\; ~\mathcal{DP}(\mathbf{x},~ \mathbf{w})~\in ~Q_{i-1} \,.$$ 
 \end{itemize}
\textbf{Inversion}. In this methodology, the goal is to compute $\mathbf{x}\in \fqn$ from given $
\mathbf{y}\in \fqm$ such that $\mathbf{y}=\p(\mathbf{x})$. The knowledge of nested sequences of input and output subspaces is used in this computation. At first glance, for $l$-layer Rainbow, the value of the unknown $\mathbf{x}$ can be represented as $\mathbf{v}+\mathbf{o_1}+\cdots+\mathbf{o}_l$ where all of the $\mathbf{o_i}\in O_i$. Fix $\mathbf{v}\in_U \mathbb{F}_q^n$.
  Then $\p$ is used in conjunction with the $i$th-layer's output subspace $Q_i$ to calculate $\mathbf{o}_i$.  For the sake of clarity, let's define the \textit{quotient space} $\overline{O}_i := O_i/O_{i+1}$.
 
 Using the knowledge of sequences of subspaces, the goal is to find $\textbf{o}_i$ for all $i$. This will lead to computing the preimage of any element from $\fqn$.  For computing $\overline{\mathbf{o}}_i\in \overline{O}_i$, use the following relation (note that, from definition, $\mathcal{P}\ (\overline{\mathbf{o}}_i)=0$),
 \begin{align*} 
&\mathcal{P}(\mathbf{v}+\overline{\mathbf{o}}_i)\ + \ Q_i \ =\ \mathbf{y}+Q_i  \\ 
\implies &\mathcal{P}(\ \mathbf{v})\ +\ \mathcal{P}\ (\overline{\mathbf{o}}_i)\ +\ \mathcal{DP}\ (\mathbf{v}, \ \overline{\mathbf{o}}_i)\ + \ Q_i \,=\ \mathbf{y}+Q_i \,.
\end{align*}
Earlier $\textbf{v}$ is fixed, so the quadratic system reduces to a linear system. The number of constraints and variables are the same for the linear system. This implies that a unique solution can be obtained with probability $(1-\frac{1}{q})$. Repeatedly running this procedure, one can compute all $\textbf{o}_i$, which implies that preimage $\mathbf{x}$ will be computed. 

In 2022, Beullens~\cite{beullens2022breaking} reduced the security level of Rainbow. He showed for small $n-m$, recovering all subspaces is significantly efficient. Also, the small finite field size accelerates the attack.
%The signature algorithm is the same as earlier. The secret key is sequences of input and output subspaces instead of secret maps $\s,\ct,\f$. In a similar way, a verifier can verify the signature. 
%\vspace{-10pt}
\subsection{Concurrent Proposals} The NIST additional signature submission call \cite{nistadditional} received a total of eleven multivariate signature schemes e.g. Mayo~\cite{beullens2022mayo}, QR-UOV~\cite{furue2023qr}, TUOV~\cite{ding2023tuov}, etc. Most of them are based on the old \emph{unbalanced Oil-Vinegar} structure. For example, Mayo~\cite{beullens2022mayo} employed a UOV structure along with a new \emph{whipped-up MQ} (WMQ) approach. QR-UOV is another variant of UOV where the public key is represented by block matrices, with each element corresponding to an element in a quotient ring~\cite{furue2023qr}. Also, in 2022, a new proposal, called IPRainbow~\cite{cartor2022iprainbow} was made by perturbing the central polynomials of the second layer by $s$ variables. This change although decreases the attack probability by $1/q^s$, the running time significantly increases due to the usage of Gr\"{o}bner basis technique for inversion.
%\subsection{IPRainbow}
%IPRainbow is a multi-layer construction based on the ground layer UOV. The signature generation and verification phase is the same as Rainbow, the only difference is in the central polynomial. 
%Here, central polynomials of the second layer are perturbed by $s$ variables, which decreases the probability of guessing a variable in $O_2$ by $1/q^s$. 
%However, running time is significantly increased during signature generation due to the presence of Gr\"{o}bner basis technique in the inversion.
%\vspace{-5pt}
\subsection{Hardness of Multivariate Cryptography}
Here, we describe other approaches used in the cryptanalysis of multivariate signatures apart from the direct solution of MQ equations.
\begin{enumerate}
%\item {\bf MQ.} Knowing the public polynomial map $\mathcal{P}$ and $\mathbf{y}=\mathcal{P}(\mathbf{x})$, the task is to find $\mathbf{x}$. This problem is called the MQ problem (multivariate quadratic), and it is known to be NP-hard \cite{johnson1979computers}.  
 %Beullens leveraged the multivariate quadratic system's special structure for a direct attack, leading to secret information recovery.
\item {\bf Min-rank.} Let $M_1,\;M_2,\;\cdots, \;M_k\;\in\;\mathbb{F}_q^{n\times m}$ be the given matrices and $r\in \mathbb{N}$, find a non-trivial linear combination (with $m_1,m_2,\cdots,m_k\in\mathbb{F}_q$) so that $\mbox{rank }(\,\sum_{i=1}^k \, m_iM_i\,)\leq \;r.$
This problem is called the \emph{min-rank }problem and has been shown to be NP-hard~\cite{buss1999computational}. The min-rank problem appeared as a cryptanalytic tool in multivariate cryptography \cite{kipnis1999cryptanalysis,faugere2008cryptanalysis,bardet2002algebraic,beullens2021improved}. This attack helps to find a linear combination of public matrices which sums up to a low-rank matrix.
\item {\bf EIP.} Find an equivalent composition of $\mathcal{P}=\mathcal{S'}\circ\mathcal{F'}\circ\mathcal{T'}$, where $\mathcal{S'}\mbox{ and }\mathcal{T'}$ are equivalent affine maps, and $\mathcal{F'}$ is an equivalent central map. The above problem is the \textit{Extended Isomorphism of Polynomials (EIP)} problem. No such hardness classification is known (though it subsumes graph isomorphism problem \cite{agrawal2005automorphisms,agrawal2006equivalence}), but for some instances, polynomial time algorithms exist \cite{kipnis1998cryptanalysis}.
 %Later researchers use the special structure of the quadratic system and improves the state of the art, like, band-separation attack \cite{ding2008new,thomae2012generalization,smith2020rainbow}, intersection attack \cite{beullens2021improved}, and most famous simple attack \cite{beullens2022breaking}.
\end{enumerate} 
%\vspace{-5pt}
\section{Our Proposal: VDOO Signature Scheme}\label{sec:scheme}
%\vspace{-5pt}
%Our scheme combines diagonals with the Oil-Vinegar in an efficient manner.
In our scheme, we introduce a new design element called \emph{diagonals} into the Oil-Vinegar scheme. %Let, $[n]$ be the set of indices of variables, we pick the first $v$ variables as vinegar variables. 
Let, $\mathbf{x}\in\fqn$,  we pick the first $v$ variables as vinegar variables. We denote the next $d$ variables as diagonal variables.  In this layer, we introduce $d$ quadratic equations. In any $i$-th ($1\leq i\leq d$) equation,  only $v+i$-th variable is unknown among $v+i$ variables. In the following layers, we apply the Oil-Vinegar technique. This means we can generate $o_1$ OV polynomials using $v+d$-vinegar variables and newly added $o_1$-oil variables. Further, we construct $o_2$ OV polynomials using $v+d+o_1$-vinegar variables and newly added $o_2$-oil variables. Finally, we have a quadratic system with $n=v+d+o_1+o_2$ variables and $m=d+o_1+o_2$ homogeneous quadratic equations.
%\vspace{-5pt}
\subsection{{\sf VDOOSetUp}: Generate Parameters}
%\vspace{-10pt}
To construct polynomial maps we need to define parameters associated with this. In this phase algorithm takes input the security parameter $\lambda$ and output the parameter tuple, that is $\mathsf{params}=(\; q,\ v,~d,~o_1,~o_2~ )\leftarrow\mathsf{VDOOSetUp}(1^\lambda)$. Here,
\begin{itemize}
 \item Finite field $\mathbb{F}_q$ which has $q$ elements.
    \item Positive integers $v,\ d,\ o_1,$ and $o_2$, where $v$ denotes the number of vinegar variables, $d$ is the number of diagonal variables, $o_1$ and $o_2$ stands for the number of first and second layer oil variables respectively. Therefore, total number of variables is $n=v+d+o_1+o_2$, and number of equations is $m=d+o_1+o_2$.
    %, and $o_2$: number of second-layer oil variables.
 %   \begin{itemize} \item $v$: number of vinegar variables, that is $v=n-m$ where    $n$ is total number of variables, and $m$: number of homogeneous equations present in the quadratic system,  \item $d$: number of diagonal variables, \item $o_1$: number of first-layer oil variables   \item $o_2$: number of second-layer oil variables. \end{itemize} 
 \end{itemize}
% \ifsubmission
% \else
% To generate the parameter set, signer can use $ \mathsf{VDOOSetUp}$ algorithm \ref{setup}

% \begin{algorithm}
% \caption{$ \mathsf{VDOOSetUp}$}\label{setup}
% \begin{algorithmic}[1]
% \Require Security parameter $\lambda$
% \Ensure Parameter tuple $\mathsf{params}=(q,v,d,o_1,o_2)$
% %, where $n$: number of variables; $m$: number of homogeneous equations; $q$: size finite field; $d$: number of diagonal variables; $o_1$: number of first layer oil variables; $o_2$: number of second layer oil variables.
% \State Generate a parameter tuple corresponding to the security level $\lambda$
% \State {\bf Return} $\mathsf{params}$
% \end{algorithmic}
% \end{algorithm}
% \fi
\subsection{VDOO Central Polynomial Map and Inversion.}
Construction of central polynomial map $\f:\fqn\to\fqm$ plays an important role in the multivariate signature schemes. To the best of our knowledge, we are the first to propose a central polynomial map that involves vinegar, diagonal, and oil variables in a three-layer construction.
\begin{itemize}
    \item\textbf{Diagonal Layer.} Here, we explain the structure of any central polynomial $f_k$ for the diagonal layer $k\in [v+1:v+d]$. Each $f_k$ is defined as follows.
  %  Suppose $f_k$ denotes a quadratic polynomial in this layer and $k\in [v+1:v+d]$. The structure of the central polynomial is as follows.
   % At first, we discuss the structure of the central polynomials for the diagonal layer. Suppose $f_k$ denotes a quadratic polynomial in this layer and $k\in [v+1:v+d]$. The structure of the central polynomial is as follows.
    $$f_{k-v}(x_1,x_2,\cdots,x_n)=\sum_{i=1}^{k-1}\alpha_{i,k}^{(k)}x_ix_k +\sum_{i,j=1,i\leq j}^{k-1} \beta_{i,j}^{(k)}x_ix_j $$ 
Each coefficient $\alpha_{ij}^{(k)}$, and $\beta_{ij}^{(k)}\in_U\mathbb{F}_q$. 
The subroutine $\mathsf{DiagPoly}(q,\;k)$ is used to generate such central polynomial $f_k$ in the diagonal layer.
%We denote this method to generate the central polynomial $f_k$ in the diagonal layer as $\mathsf{DiagPoly}(q,\;k)$.
\item \textbf{First Oil Layer.}  In this oil layer, we use $v+d$ variables as vinegar variables and next $o_1$ variables as oil variables. All these variables help us to construct $o_1$ homogeneous quadratic polynomials of the following form.
%Any first-layer central polynomial has the following form,
 $$f_{k-v}(x_1,x_2,\cdots,x_n)~=~\sum_{i=1}^{v+d}~\sum_{j=1}^{v+d} ~\alpha_{ij}^{(k)}~x_ix_j+ \sum_{i=1}^{v+d}~\sum_{j=v+d+1}^{v+d+o_1}~ \beta_{ij}^{(k)}~x_ix_j $$
where $k\in[v+d+1:v+d+o_1]$, $\alpha_{ij}^{(k)}$, and $\beta_{ij}^{(k)}\in_U\mathbb{F}_q$.

\item \textbf{Second Oil Layer.} The topmost oil layer has $v+d+o_1$ vinegar and $o_2$ oil variables. That means, it has $o_2$ quadratic equations. Those equations are of the form 
 $$f_{k-v}(x_1,x_2,\cdots,x_n)~=~\sum_{i=1}^{v+d+o_1}~\sum_{j=1}^{v+d+o_1} \alpha_{ij}^{(k)}~x_ix_j~+ \sum_{i=1}^{v+d+o_1}~\sum_{j=v+d+o_1+1}^{v+d+o_1+o_2} ~\beta_{ij}^{(k)}x_ix_j ,$$
 where $k\in[v+d+o_1+1:v+d+o_1+o_2=n]$ and $\alpha_{ij}^{(k)}$, and $\beta_{ij}^{(k)}\in_U\mathbb{F}_q$. 
 We denote this as $\mathsf{OVPoly}(q,\ v,\ o)$ to generate a oil-vinegar central polynomial (according to \ref{defn:OV}) which has $v$ vinegar variables and $o$ oil variables.
\end{itemize}
%\medskip {\bf Generate  VDOO Central Polynomial.}
%The algorithm $\mathsf{VDOOCentPoly(params)}$ is used to generate VDOO central polynomial ( see algorithm \ref{centpoly}). It takes two algorithms as ingredients, $\mathsf{DiagPoly}$ and $\mathsf{OVPoly}$. 
Here, Algorithm~\ref{centpoly}, uses $\mathsf{OVPoly}$ and $\mathsf{DiagPoly}$ to generate a VDOO central map $\f$.
\begin{algorithm}
\caption{$\mathsf{VDOOCentPoly}$}\label{centpoly}
\begin{algorithmic}[1]
\Require Parameter tuple $params=(q,v,d,o_1,o_2)$
\Ensure Central map $\f=~(f_1,\cdots,f_m)~:\fqn~\to~\fqm$
\State Compute $m=d+o_1+o_2$ and $n=v+m$.
\State \textbf{for } $1\leq i\leq d$
\\ $~~~~~f_i\leftarrow ~\mathsf{DiagPoly}~(\;q,\;i)$ 
\State \textbf{for } $d+1\leq i\leq d+o_1$
\\ $~~~~~f_i\leftarrow ~\mathsf{OVPoly}~(\;q,\; v+d,\; o_1)$
\State \textbf{for } $d+o_1+1\leq i\leq m$
\\ $~~~~~f_i\leftarrow ~\mathsf{OVPoly}~(\;q,\; v+d+o_1,\;o_2)$ 
\State {\bf Return} VDOO central polynomial $\f$
\end{algorithmic}
\end{algorithm}

\medskip {\bf Inversion}. The main computational bottleneck of UOV-based constructions is the inversion of the central polynomial. It requires Gaussian elimination which runs in $O(N^3)$.
%Time complexity of this part of computation is around $O(N^3)$. 
However, in our scenario inversion of the diagonal polynomials is straightforward as there is only one unknown variable.
%variable is unknown and rest of them are known.  
Nevertheless, the inversion of OV polynomials in the remaining two layers each needs a Gaussian elimination. Therefore, inverting VDOO central polynomial map needs two Gaussian elimination only. This is shown in Algorithm~\ref{centpolyinversion}. Following two algorithms help to compute the inverse of the VDOO central polynomial.
%We present the inversion of VDOO central polynomial map in an algorithmic way (see \ref{centpolyinversion}). 
\begin{description}
    \item[$\mathsf{ST}$] The subroutine {\it Substitution} or $\mathsf{ST}$ converts a bunch of oil-vinegar polynomials to a bunch of linear polynomials consists of oil variables by fixing the vinegar variables. That means, $\mathsf{ST}$ substitutes vinegar variables $x_1,\cdots , x_{v}$ by random values (in $\fq$) in the bunch of $o$ oil-vinegar polynomials $(f_i)_{i=1}^o$ and converts it to a bunch of linear polynomials of $o$ oil variables $(\Tilde{f}_i)_{i=1}^o$.
    \item[$\mathsf{GE}$] The $\mathsf{GE}_{(q,l)}$ denotes Gaussian elimination for $l$ unknowns over the linear system of equations $(~\Tilde{f}_i=y_i~)_{i=1}^l$ over $\fq$. It returns a failure when the rank of the matrix representing the linear system is less than $l$.
\end{description}
%Here, the subroutine $\mathsf{ST}$ fixes $x_1,\cdots , x_{l}$ in $f_i$ and convert it to $\Tilde{f}_i$ for all $i.$ Here, we use two subroutines $\mathsf{ST}$ and $\mathsf{GE}$. The $\mathsf{ST}$ algorithm substitutes $x_1,\cdots , x_{l}$ values in quadratic $f_{1},\cdots,f_{l}$ polynomials and return a linear pol $(\Tilde{f}_{1},\cdots,\Tilde{f}_{l})$. 
%And, the algorithm $(x_1,\cdots,~x_l)~\leftarrow~\mathsf{GE}_{(q,l)}(~\Tilde{f}_{1}=y_1,~\cdots~,~\Tilde{f}_{l}=y_l~)$ denotes the computation of Gaussian elimination over the $l$ unknowns linear system $(~\Tilde{f}_i=y_i~)_{i=1}^l$. The $\mathsf{GE}_{(q,l)}$ denotes Gaussian elimination for $l$ unknowns over the linear system of equations $(~\Tilde{f}_i=y_i~)_{i=1}^l$. The function $\mathsf{GE}$ returns a failure when the rank of the matrix representing the linear system is less than $l$.
\begin{algorithm}
\caption{$ \mathsf{VDOOCentPoly\_Inversion}$}\label{centpolyinversion}
\begin{algorithmic}[1]
\Require Central map: $\mathcal{F}=~(f_1,\cdots, f_m)~:\mathbb{F}_q^n~\to~\mathbb{F}_q^m$ and $\mathbf{y}\in \mathbb{F}_q^m$, and $\mathsf{params}$.
\Ensure A vector $\mathbf{x}\in \mathbb{F}_q^n$ such that $\f(\mathbf{x})=\mathbf{y}$.
\State $m\leftarrow d+o_1+o_2$ and $n\leftarrow v+m$
\State Randomly fix first $v$-vinegar variables $x_1,\cdots,x_v~{\leftarrow}_\$~ \mathbb{F}_q$
\State \textbf{for } $1\leq i\leq d$\\
$~~~~~~$compute $x_{v+i}$ using $y_i,\;x_1,\; \cdots,\;x_{v+i-1}$ and $f_i$.
\State $(~\Tilde{f}_{d+1},\cdots,
\Tilde{f}_{v+d})\leftarrow \mathsf{ST}\big(f_{d+1} ( x_1,\cdots , x_{v+d})~,\cdots,~f_{d+o_1}(x_1,\cdots , x_{v+d})\big)$ 
\State $(x_{v+d+1},\cdots,x_{v+d+o_1})\leftarrow \mathsf{GE}_{(q,o_1)}(\Tilde{f}_{d+1}=y_{d+1},\cdots, \Tilde{f}_{d+o_1}=y_{d+o_1})$.
\State $(\Tilde{f}_{d+o_1+1},\cdots,
\Tilde{f}_{m})\leftarrow \mathsf{ST}\big(f_{d+o_1+1} ( x_1,\cdots , x_{n-o_2}),\cdots,f_{m}(x_1,\cdots , x_{n-o_2})\big)$ 

%\State Substitute $x_1,\cdots , x_{v+d+o_1}$ values in $f_{d+o_1+1},\cdots,f_{m}$ polynomials.
\State $ (x_{v+d+o_1+1},\cdots,x_{n})\leftarrow\mathsf{GE}_{(q,o_2)}(\Tilde{f}_{d+o_1+1}=y_{d+o_1+1},\cdots, \Tilde{f}_{m}=y_m)$ 
\State {\bf Return} $\mathbf{x}\in \mathbb{F}_q^n$
\end{algorithmic}
\end{algorithm}

\subsection{{\sf VDOOKeyGen}: VDOO Key Generation}
The $\mathsf{VDOOKeyGen}$ in Algorithm~\ref{algokeygen} generates two random invertible affine maps $\s:\fqm\to\fqm$ and  $\ct:\fqn\to\fqn$ along with the VDOO-central map $\f:\fqn\to\fqm$. %The private key is the individual information these three maps. 
Here, {\it secret/signing key} is $\s,~\f$, and $\ct$ and {\it public/verification key} is the composition map $\p$, where $\p=\s\circ \f\circ \ct:\fqn\to\fqm  $. Note that, the individual information of secret maps allows user to compute the inverse of $\p$ efficiently. We denote $S\leftarrow \mathsf{randomMatrix}\;(q,\;m,\;seed)$ to generate a random $m\times m$ matrix over $\mathbb{F}_q$ from a $seed$, $\mathsf{invMat}\;(q,\;m,\;S)$ helps to compute the inverse of a $m\times m $ matrix $S$ over $\fq$, and $\mathsf{Affine}(S,\mathbf{a})$ computes $\s\leftarrow S\cdot \mathbf{x}+\mathbf{a}.$

\begin{algorithm}
\caption{$ \mathsf{VDOOKeyGen}$}\label{algokeygen}
\begin{algorithmic}[1]
\Require Parameter tuple $\mathsf{params}$.
\Ensure Generate public and private key pair.
\begin{itemize}
    \item Public key: $\mathsf{pk}=\p$.
    \item Secret key: $\mathsf{sk}=$  $\s$, $\ct$, and $\f$.
\end{itemize}
\State $m\leftarrow d+o_1+o_2$ and $n\leftarrow m+v$
\State $seed\leftarrow PRNG(1^\lambda)$\Comment{$\lambda$ is the security parameter}
\While {$(\det (S)\neq 0~\&\&~\det (T)\neq 0)$} \State
        $S\leftarrow\mathsf{randomMatrix}\ (q,\,m\,,seed)$ \Comment{$S\in_U\fq^{m\times m}$}\State
        $T\leftarrow\mathsf{randomMatrix}\ (q\,,n\,,seed)$ \Comment{$T\in_U\fq^{n\times n}$}
\EndWhile
\State $\mathbf{a}\in_U \fqm$ and  $\mathbf{b}\in_U\fqn$ \Comment{generate two random vector}
\State $invS\leftarrow \mathsf{invMat}(q,m,S)$ and $invT\leftarrow \mathsf{invMat}(q,n,T)$ \Comment{compute inverse of matrices}
\State $\s\leftarrow \mathsf{Affine}(S,\mathbf{a})$ and  $\ct\leftarrow \mathsf{Affine}(T,\mathbf{b})$ \Comment{Constructing invertible affine maps}
\State $\f\leftarrow\mathsf{VDOOCentPoly(\mathsf{params})}$ \Comment{generate VDOO central map}
%\State Generate a $m\times m$ random matrix. Check it is invertible or not, if not then regenerate and check. Construct an affine map $\s:\fqm\rightarrow\fqm$ using the random invertible map. \State Compute $\f\leftarrow\mathsf{VDOOCentPoly(\mathsf{params})}$.\State Generate a $n\times n$ random matrix. Check it is invertible or not, if not then regenerate and check. Construct an affine map $\ct:\fqn\rightarrow\fqn$ using the random invertible map. 
\State Compute $\p\leftarrow\s\circ\f\circ\ct$
\State {\bf Return} 
$\mathsf{pk}=\p$ and $\mathsf{sk}=(invS,~\mathbf{a}$, $invT,~\mathbf{b})$ 
(equivalently sending $\s,\mbox{ and }\ct$). 
\end{algorithmic}
\end{algorithm}

\iffalse
\begin{algorithm}
\caption{$ \mathsf{VDOOKeyGen}$}\label{algokeygen}
\begin{algorithmic}[1]
\Require Parameter tuple $\mathsf{params}$.
\Ensure Generate public and private key pair.
\begin{itemize}
    \item Public key: $\p$
    \item Secret key: two random invertiable affine map $\s$, $\ct$, and VDOO central map $\f$
\end{itemize}
\State Generate a $m\times m$ random matrix. Check it is invertible or not, if not then regenerate and check. Construct an affine map $\s:\fqm\to\fqm$ using the random invertible map. 
\State Compute $\f\leftarrow\mathsf{VDOOCentPoly(\mathsf{params})}$.
\State Generate a $n\times n$ random matrix. Check if it is invertible or not, if not then regenerate and check. Construct an affine map $\ct:\fqn\to\fqn$ using the random invertible map. 
\State Compute $\p\leftarrow\s\circ\f\circ\ct$
\State {\bf Return} $\p$ as public key and private key $\s$, $\ct$ and $\f$.
\end{algorithmic}
\end{algorithm}\fi
\subsection{{\sf VDOOSign}: VDOO Signature Generation}
Similar to the other OV based constructions~\cite{beullens2022mayo,ding2023tuov,kipnis1999unbalanced,ding2005rainbow}, we use the hash-and-sign paradigm for our signature algorithm as shown in Algorithm~\ref{vdoosign}.
%VDOO signature scheme belongs to hash and sign paradigm. In this phase, we need a hash function as an ingredient. This hash function maps an arbitrary length message to an element of $\fqm$. 
We use a hash function $\mathcal{H}:\{0,1\}^*\to \mathbb{F}_q^m$. 
Signer knows each polynomial map, so it can compute the inverse of each map \textit{i.e.} $\s^{-1}$, $\f^{-1}$, and $\ct^{-1}$.
%To compute the inverse chain of mappings, we first, compute $\s^{-1}$, then the inverse of the central map $\f^{-1}$ and later $\ct^{-1}$. We also compute $\s^{-1}$, $\f^{-1}$, $\ct^{-1}$. 
If $\mathsf{GE}$ reports a failure during the computation of $\f^{-1}$, we 
restart the process by regenerating the salt and repeating the entire procedure. Finally, the signature is computed as $\p^{-1}(\mathcal{H}(\mathcal{H}(msg)||salt))$.
\begin{algorithm}
\caption{$ \mathsf{VDOOSign}$}\label{vdoosign}
\begin{algorithmic}[1]
\Require $\mathsf{sk}=(invS,~\mathbf{a}$, $invT,~\mathbf{b})$, message $msg$, and $\mathcal{H}:\{0,1\}^*\to\; \mathbb{F}_q^m$
\Ensure a signature $\sigma=(\mathbf{s},salt)$
\State $salt\longleftarrow PRNG$
\State Use hash function $\mathbf{d}\leftarrow\mathcal{H}(\mathcal{H}(msg)||salt)$
\State Compute $\mathbf{t}=invS\times (\mathbf{d}-\mathbf{a})$\Comment{$\mathbf{t}=\s^{-1}(\mathbf{d})$}
\State Compute $\mathbf{y}=\f^{-1}(\mathbf{t})$ using $\mathsf{VDOOCentPoly\_Inversion}$ \ref{centpolyinversion}.
\State Compute $\mathbf{s}=invT\times (\mathbf{y}-\mathbf{b})$\Comment{$\mathbf{s}=\ct^{-1}(\mathbf{y})$}
\State {\bf Return} signature $\sigma=(\mathbf{s},salt)$
\end{algorithmic}
\end{algorithm}

%{\bf Computational bottleneck.} The major computational overhead within the signature generation process is the Gaussian elimination, which carries a complexity of $O(N^3)$. While it might appear that computing $\s^{-1}$ and $\ct^{-1}$ also incur substantial costs (step two and three), it's important to note that these steps we efficiently executed during the one-time key generation process. As a result, these particular steps do not add any additional cost overhead.
\noindent
{\bf Efficiency analysis.} As mentioned earlier, the major computational overhead of OV-based schemes is the Gaussian elimination procedure. In VDOO, during signing, we have to compute only one Gaussian elimination \textit{i.e.} computation of $\f^{-1}$. The computation of $\s^{-1}$ and $\ct^{-1}$ can be done during the key-generation procedure. In VDOO the computation of $\f^{-1}$ is also very efficient compared to other OV-based schemes as the number of unknowns is smaller in VDOO as shown in Table~\ref{tab:compmul}. 
\subsection{{\sf VDOOVerif}: VDOO Verification}
%Our verification procedure is very simple. It is just a polynomial evaluation of MQ map. This requires  $O(N^3)$-field operations. 
Our verification procedure is simple. It needs a polynomial evaluation of $\p$, requiring just $O(N^3)$ field operations. Compute $\mathbf{d}'= \p(\mathbf{s})$ from public key $\p$ and signature $\sigma=(\mathbf{s},\;salt)$
%Given a public key $\p$ and signature $\sigma=(\mathbf{s},\;salt)$, it computes $\mathbf{d}'\leftarrow \p(\mathbf{s})$. 
The signatures is accepted if $\mathbf{d}'=\mathcal{H}(\mathcal{H}(msg)~||~salt)$ holds, else rejected.
%Then compares it with $\mathbf{t}=\mathcal{H}(msg)$. If $\mathbf{t}=\mathbf{t}'$ matches, then report accept else reject.
\begin{algorithm}
\caption{$ \mathsf{VDOOVerif}$}\label{verf}
\begin{algorithmic}[1]
\Require  $\mathsf{pk}=\p$; message $msg$; signature $\sigma=(\mathbf{s},salt)$ and $\mathcal{H}:\{0,1\}^*\to \mathbb{F}_q^m$.
\Ensure  {\it accept} or {\it reject}
%\State $salt\leftarrow PRNG$
\State Use hash function to compute  $\mathbf{d}\leftarrow \mathcal{H}(\mathcal{H}(msg)~||~salt)$
\State Compute $\mathbf{d}'=\p(\mathbf{s})$
\If{$\mathbf{d}=\mathbf{d}'$} output {\it accept}
\Else{}  {\it reject}
\EndIf
\State {\bf Return } {\it accept } or {\it reject}
\end{algorithmic}
\end{algorithm}%\vspace{-3em}
\subsection{Key Size Computation}
Our VDOO contains one diagonal layer and two UOV layers. The size of the private key is determined first, followed by the size of the public key.
\begin{itemize}
  \item Size of the central map $\mathcal{F} $ for a diagonal layer having $d$-diagonal polynomials is $\sum_{i=1}^d\left(\dfrac{v_i(v_i+1)}{2}+v_i\right)$ field elements. The first diagonal layer has $v_1=n-m$ vinegar variables. In any diagonal layer, a central polynomial $f_i$ has $v_i$ vinegar variables and $f_{i+1}$-th polynomial has $v_{i+1}=v_i+1$ vinegar variables.
    \item Size of the central map $\mathcal{F} $ for a UOV layer is  around   $o\times \left(\dfrac{v(v+1)}{2}+ov\right)$ field elements. Such UOV layer has $v$ vinegar variables and $o$ oil variables.
  
\end{itemize}
 The sizes of the two affine transformations are as follows: for $\mathcal{S}$ we need $m(m+1)$, while for $\mathcal{T}$ we need $n(n+1)$, field elements. These maps can be generated using a random seed.
%However, we do not need to include the size of $\mathcal{S, T}$ as this does not need to be counted in the public key.

Now we are interested in computing the size of the public key of standard VDOO. Each $n$-variate quadratic polynomial requires $ \frac{(n+1)(n+2)}{2}$ field elements. Therefore, the size of the public key is $m\frac{(n+1)(n+2)}{2}$. Further optimization of public key is possible \cite{petzoldt2010cyclicrainbow,petzoldt2010selecting}.
%Just like the reductions used in Petzoldt \textit{et al.} \cite{petzoldt2010selecting} and cyclicRainbow \cite{petzoldt2010cyclicrainbow}, we can also enhance the performance and significantly reduce the size of the public key in VDOO.
It optimized the public key size from $O(mn^2\log q)$ to $O(m^3\log q)$.
\subsection{Subspace Description of VDOO Central Polynomial}
\begin{figure}[h]
\centering
\includegraphics[scale=.6]{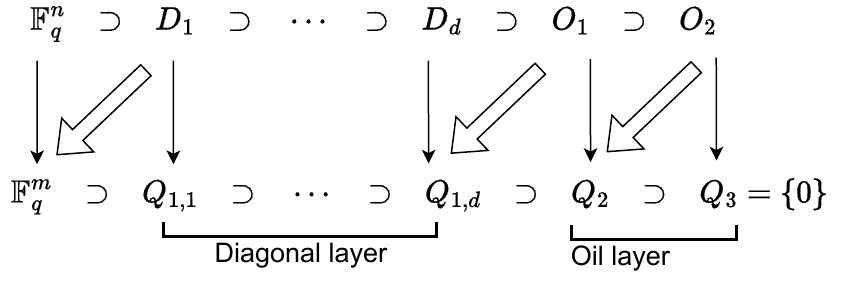}
\caption{Central polynomial of VDOO}    \label{fig: vdoo}
\end{figure}

Our scheme can be explained through Beullens's subspace descriptions~\cite{beullens2021improved}. This description is useful to understand the cryptanalysis of VDOO. In this case, we have $d+2$ input and output subspaces. These sequences of nested subspaces are as follows. 
\begin{itemize}
    \item \textbf{Input subspaces } 
    $\fqn\supset D_{1}\;\supset \;D_{2}\;\supset\;\cdots\;\supset \;D_{d}\;\supset \;O_1\;\supset \;O_2 \,.$
    
    \item \textbf{Output subspaces } 
    $\fqm\supset Q_{1,1}\supset Q_{1,2}\supset\cdots\supset Q_{1,d}~\supset Q_2\supset \,Q_3=~\{0\} \,.$
\end{itemize}
In the Figure~\ref{fig: vdoo} (single arrow denotes $\p$ and bold arrow denotes $\mathcal{DP}(\mathbf{x},\cdot)$), these following relations will hold: $\dim(D_{i})=\dim(D_{i+1})+1 $ and $\dim(Q_{1,i})=\;\dim(Q_{1,i+1})+1 $ for $1\leq i< d$. Also, $\dim(D_{1})=m$, $\dim(D_{i})=\dim(Q_{1,i-1}) $   for $1<
 i\leq d$. In addition, $\dim (O_1)=\dim(Q_{1,d})=o_1+o_2$, $\dim (O_2)=\dim(Q_2)=o_2$.
% \begin{figure}[h]
% \centering
% \includegraphics[width=13cm, height=5cm ]{vdoo.pdf}
% \caption{Central polynomial of VDOO}    \label{fig: vdoo}
% \end{figure}

The signer first fixes $\textbf{v}\in_U\fqn$. Since $\dim(\Tilde{D}_{i})= \dim (D_{i})- \dim (D_{i+1})=1$, so for diagonal layer computing $ \textbf{d}_{1}\ ,\cdots\ , \textbf{d}_{d}$ is very easy. Once these vectors are found, then update $\textbf{v}\leftarrow \textbf{v}+\textbf{d}_{1}+\cdots+\textbf{d}_{d}$. Now, signer needs to solve for $ \Tilde{\textbf{o}}_1 \in \Tilde{O_1}(=O_1/O_2)$, so that the following relation holds. Note that, $\dim(\Tilde{O_1})=o_1$.
$$\p(\textbf{v})+~\mathcal{DP}(\textbf{v},\Tilde{\mathbf{o}_1})  =\textbf{t}\mod~ Q_2.$$
We know that the above equation is a linear system of $o_1$ variables and $o_1$ equations. With the probability $(1-1/q)$, the signer will able to compute $\mathbf{o}_1$. Then signer again updates $\mathbf{v}\leftarrow\mathbf{v}+\mathbf{o}_1$ and follow a similar strategy to find $\textbf{o}_2\in O_2$. Thus the signer can finally compute the pre-image of $\mathbf{t}$.
\section{Security Analysis of VDOO}\label{sec:sec_analysis_params}
%The hardness of multivariate cryptography is based on three hard problems, multivariate quadratic problem, min-rank, and extended isomorphism problem (EIP). All cryptanalysis of multivariate schemes uses best known algorithms to solve above problems. 
Cryptanalysis that targets solving the MQ problem directly, is known as the direct attack in multivariate cryptography~\cite{bardet2002algebraic,faugere1999new,faugere2002new,bettale2009hybrid}.
Later researchers have used the special structure of the quadratic system and improved the state-of-the-art, like, band-separation attack \cite{ding2008new,thomae2012generalization,smith2020rainbow}, intersection attack \cite{beullens2021improved}, and simple attack \cite{beullens2022breaking}.

%Beullens capitalized on the inherent structure of Rainbow's multivariate quadratic systems and employed a direct attack  method to retrieve secret key of \cite{beullens2022breaking}.
%\vspace{-2mm}

%Next we study all celebrated cryptanalysis results from popular schemes Rainbow and OV. Because our scheme uses the construction of these schemes in a beautiful way. These cryptanalysis are playing a vital role to fixed parameter sets for VDOO. Due to the inclusion of Rainbow in the NIST competition, researchers focused on the cryptanalysis of Rainbow from the last decade. Many interesting results came out as output of the efforts, like direct attack \cite{bardet2002algebraic,faugere1999new,faugere2002new}, min-rank attack \cite{billet2006cryptanalysis,bardet2002algebraic,bardet2020improvements,baena2022improving}, band-separation attack \cite{ding2008new,thomae2012generalization,smith2020rainbow}, rectangular min-rank and intersection attack \cite{beullens2021improved}, and most famous simple attack \cite{beullens2022breaking}. Most  of the attacks try to find a vector in the top oil layer, and then using this information the attacker is able to learn the entire subspace. Finding this vector either requires to use a direct attack or a min-rank algorithm. 
To determine the complexity of the attacks described below by the number of field multiplications required to perform the attack. One $\fq$-field multiplication needs $(2(\log_2 q)^2+\log_2 q)$ gates. 
%We follow the following relation between the number of gates and the number of field multiplications, $$\# \mbox{ gates }=\#\mbox{ field multiplications }\cdot \Big(2(\log_2 q)^2+\log_2 q\Big).$$
Here, each $2(\log_2 q)^2$-bit stands for one $(\log_2 q)^2$-bit multiplication (represented as AND gates) and the same number of additions (represented as XOR gates) during one $\fq$-multiplication. Additionally, $\log_2 q$ bits are needed for $\log_2 q$-bit additions involved in one $\fq$-addition, which is required for each field multiplication that occurs during an attack. For example, the cost one $\mathbb{F}_{16}$-multiplication requires 36 gates. Such a strategy to determine the complexity is standard and has been also followed in other MQ-based signature schemes~\cite{furue2023qr,beullens2022mayo,faugereprov}.

Henceforth, in this document, we use the parameter set $(q,v,d,o_1,o_2)=(16, 60, 30,$ $ 34, 36)$ as an example to demonstrate the complexity of the following attacks. Incidentally, this is also our SL-1 parameter. Our full parameter set is given in Table.~\ref{tab:vdoo}.
%We set $\mathsf{params}=(q,v,d,o_1,o_2)=(16, 60, 30, 34, 36)$ as security level one (SL-1)  parameter set, as specified in \cite{chen2017nist}. Subsequently, we analyze the attack complexity against this parameter set for state-of-the-art attacks.
%\vspace{-3pt}
\subsection{Direct Attack on VDOO}
The direct attack is the fundamental methodology for forging any multivariate signature scheme. To counterfeit a VDOO signature, an attacker aims to solve an underdetermined system with $n$ variables and $m$ homogeneous equations ($n>m$), to find $\mathbf{s}$ such that $\p(\mathbf{s})=\mathbf{t}$. The basic approach involves converting this underdetermined system into a determined one by fixing $n-m$ variables. Subsequently, quadratic system-solving techniques like the Wiedemann XL algorithm~\cite{wiedemann1986solving,courtois2000efficient} or Gröbner basis methods such as F4 or F5~\cite{faugere1999new,faugere2002new} are applied. %The primary strategy involves transforming this under-determined system into a determined system by fixing $n - m$ variables. Subsequently, quadratic system solving techniques like the Wiedemann XL algorithm \cite{wiedemann1986solving,courtois2000efficient}, or Gr\"{o}bner basis methods such as F4 or F5 \cite{faugere1999new,faugere2002new} are employed. 
Another approach named hybrid approach~\cite{bettale2009hybrid} involves guessing $k$ variables prior to solving the system. The time complexity of this attack, using the approach outlined in \cite{bettale2009hybrid}, is expressed in terms of field multiplications as: $$\min_{0\leq k\leq m} q^k\cdot 3\cdot\binom{m-k+d}{d}^2\binom{m-k}{2}$$  Here, $k$ denotes the number of variables fixed during the algorithm, and $d$ represents the smallest integer for which the coefficient of $t^d$ in the series $\frac{(1-t^2)^m}{(1-t)^{m-k}}$ is non-positive. 

 %This attack is the most fundamental attack for any multivariate scheme. To forge VDOO signature, attacker tries to solve an under-determined system that has $n$-variables and $m$-homogeneous equation ($n>m$), that is, find $\mathbf{s}$ such that$\p(\mathbf{s})=\mathbf{t}$ . The most fundamental approach is to convert this under-determined system to a determined system by fixing $n-m$ variables. Then apply quadratic system solving techniques such as  Wiedemann XL algorithm \cite{wiedemann1986solving,courtois2000efficient}, Gr\"{o}bner basis method like as F4 or F5 \cite{faugere1999new,faugere2002new}. Another way is to guess $k$-variables before solving the system. This approach is known as hybrid approach \cite{bettale2009hybrid}. The time complexity of this attack, using \cite{bettale2009hybrid} approach is (in terms of field multiplications as): $$\min_{0\leq k\leq m} q^k\cdot 3\binom{m-k+d}{d}^2\binom{m-k}{2}$$ where $k$ is the number of variables to be fixed during the algorithm and $d$ is the smallest integer for which the coefficient of $t^d$ in the series $(1-t^2)^m/(1-t)^{m-k}$ is non-positive. Quantum algorithms rely on Grover's search \cite{grover1996fast} to reduce the search space, that is the number of field multiplications is reduced by a factor of $q^{k/2}$.  

\smallskip \noindent{\bf \textit{Example for SL-1 parameters}.} Our level one parameter set has 160 variables and 100 constraints. According to \cite{bettale2009hybrid}, we fix 60 variables. Now in the algorithm, if we fix twelve variables, then the value of $d$ is 28. The total complexity is around $2^{280}$. 
%\vspace{-1em}
\subsection{\bf Simple Attack on VDOO} 
In 2022, Beullens proposed the \textit{simple attack} against Rainbow~\cite{ding2005rainbow}. For Rainbow, this highly effective attack reduces $n$-unknown and $m$-constraints in the quadratic system to $n-m$-unknown and $m$-constraints. Now an attacker can apply the same methodology on VDOO to recover the secret key. Recall from Figure.~\ref{fig: vdoo}, $\p$ is the public polynomial map, and sequences of nested input and output subspaces are,

\begin{itemize}
    \item {\bf \textit{Input subspaces}}
   $ \fqn\;\supset D_{1}\;\supset D_{2}\;\supset\;\cdots\;\supset D_{d}\;\supset O_1\;\supset \;O_2 \,.$
    \item {\bf \textit{Output subspaces}}
    $\fqm\;\supset ~Q_{1,1}\supset \;Q_{1,2}\;\supset\;\cdots\supset \;Q_{1,d}\;\supset Q_2\;\supset \;Q_3\;=\;\{0\} \,.$
\end{itemize}

The main crux of the simple attack lies in finding a vector within $O_2$ (as depicted in Figure.~\ref{fig: vdoo}). To achieve this, the attacker must solve a quadratic system with $n-m$ unknowns and $m$ constraints using the XL algorithm. This computational step constitutes the most significant component of the entire attack. Here is a step-by-step outline detailing the cryptanalysis of our scheme using the simple attack.
\begin{description}
    \item[Input:] Public polynomial map $\p$.
    \item[Output:] Recover sequences of subspaces.
    \item[Find a vector $\textbf{o}\in O_2$:] 
    Choose $\textbf{v}\in_U\fqn$. Then from Figure.~\ref{fig: vdoo},  $\dpol_\textbf{v}~:~\fqn\rightarrow\fqm$ is a linear map, in particular it maps $O_2$ to $Q_2$. The attacker uses this linear relation to reduce the number of unknowns present in the quadratic system. Therefore, to find a vector, an attacker should solve the following system. 
     \begin{align*}
    \dpol_\textbf{v}(\textbf{o})&=0\\
    \p(\textbf{o})&=0
    \end{align*}
With probability $\approx1/q$, the attacker successfully guesses a vector in $O_2$.  Later, the attacker deploys the XL algorithm to solve the quadratic system of $n-m$-unknowns and $m$-constraints. Thus attacker recovers $\mathbf{o}$.
\item [Recover $Q_2$:] Attacker will retrieve $Q_2$ using the information $\textbf{o}\in O_2$. Note that, $\dpol_{\mathbf{o}}:O_2\rightarrow Q_2$ is a linear map. Therefore,
    $$\mbox{Span}\{~ \dpol_{\textbf{o}}(\textbf{e}_1),~\cdots,  ~\dpol_{\textbf{o}}(\textbf{e}_n)~\} \subseteq Q_1$$
    for some linearly independent vectors $\mathbf{e}_i$. For enough such $\textbf{e}_i$'s equality will hold.
    
    \item[Recover $O_2$:]  To recover $O_2$, solve the following system of linear equations. Because with high probability kernel of $\dpol_{\mathbf{o}}$ matches with $O_2$.
    \begin{align*}
        \dpol_\textbf{o}(\textbf{e}_1)&\equiv0\mod ~Q_2\\
        \dpol_\textbf{o}(\textbf{e}_2)&\equiv0\mod ~Q_2\\
        \vdots\\
        \dpol_\textbf{o}(\textbf{e}_n)&\equiv0\mod ~Q_2
    \end{align*}
\item [Recover a vector $\mathbf{o}'\in O_1$:] Now the quadratic system $\p$ reduces to $m'=m-o_2$ equations and $n'=n-o_2$ variables. To recover $O_1$, the goal of the attacker is to find a vector in $\mathbf{o}\in O_1$. Again attacker will guess a vector $\mathbf{v}'\in \mathbb{F}_q^{n'}$. Like above, a similar argument shows that $\dpol_{\mathbf{v}'}:~O_1\rightarrow Q_{1,1}$ is a linear map and the attacker tries to solve the following systems mod $Q_2$. 
\begin{align*}
    \dpol_{\textbf{v}'}(\textbf{o}')&=0 \mod ~Q_2\\
    \p(\textbf{o}')&=0 \mod ~Q_2
    \end{align*}
    The attacker runs the XL algorithm to solve the quadratic system of $n'-m'$-unknowns and $m'$-constraints.
    \item [Recover $O_1$:] Attacks follows same approach as recovering $O_2$ to recover $O_1$. Here, an attacker solves a system $\dpol_{\mathbf{o}'}(\mathbf{e}_i')\equiv 0\mod ~Q_1$ for $i\leq n'.$
    \item[Recovering vectors from diagonal layer:]
%  At this moment, the attacker is able to remove the top layer $O_2$. So (s)he reduces the earlier system to $m_1=m-o_2$-constraints in $ n_1=n-o_2$ unknowns. Attacker again applies the first step of this algorithm to find a vector in $O_1$, which means (s)he again needs to solve a quadratic system of $n_1-m_1$-unknowns and $m_1$ constraints. Once a vector in $O_1$ is found, then step-2 helps to recover the entire $O_1$  and $Q_1$. 
  The only task that remains is to find all the diagonal vectors. The attacker can apply Wolf et al.'s~\cite{wolf2006security} trick to find all the diagonal vectors in the layer. Here observe that the computation of finding a vector in $O_2$, dominates the computation of finding a vector in $O_1$. \end{description}

   \textbf{Attack Complexity.} The complexity of the first steps dominates the complexity of other steps involved in this algorithm. Basically, a system of $n$ variables and $m$ non-linear equations reduces to a system of $m$ homogeneous equations with $n-m $ variables. This computation can be performed via XL algorithm and it requires
    $$3\cdot q\binom{n-m-1+d}{d}^2\binom{n-m-1}{2}$$ field operations, 
    where $d$ is the operating degree of the algorithm. It means, $d$-is the smallest positive integer so that the coefficient of $t^d$ in the power series $(1-t^2)^m/(1-t)^{n-m}$ is non-positive. 

{\bf \textit{Example for SL-1 parameters}. }
%For the security level one parameter set of $\mbox{VDOO}$ \textit{i.e.} $(q,v,d,o_1,o_2)=(16, 60,30,34,36)$. 
Apply Beullens’  trick to guess a vector in $O_2$, which happens with probability $1/q$. Finding one vector on $O_2$ asks to solve a quadratic system of $100$-variables 60-unknowns. This computation is the most costly in the entire algorithm. Solving this quadratic system needs $2^{134}$ field operations. The guessing needs $1/q$ search and cost of one $\mathbb{F}_{16}$- multiplication needs 36 gates. Therefore, this parameter set provides approximately at-least 128-bit security.
%\vspace{-1em}
\subsection{Rectangular Min-rank Attack on VDOO}
   Rectangular min-rank attack is proposed by Beullens  \cite{beullens2021improved}. We first describe the attack against VDOO and then compute the required attack complexity to perform this attack against VDOO.
Attacker starts with $n\times m$-rectangular matrices $M_1,\;M_2\;,\cdots,\;M_n $ over $\mathbb{F}_q$ where each $M_i$ is defined as 
$$M_i= 
 \begin{bmatrix}
~\mathcal{DP}(\mathbf{s}_1,\mathbf{s}_i) \\
~\mathcal{DP}(\mathbf{s}_2,\mathbf{s}_i) \\
\vdots \\
~\mathcal{DP}(\mathbf{s}_n,\mathbf{s}_i)
\end{bmatrix}  $$
where $(\mathbf{s}_i)_{i=1}^n$ is a basis  of $\mathbb{F}_q^n$. 

Let $\mathbf{o}_2\in\mathbb{F}_q^n$.  The bi-linearity of $\mathcal{DP}$ implies 
$$M\,:=\,\sum_{i=1}^n o_{2i} M_i \,:=\, 
 \begin{bmatrix}
~\mathcal{DP}(\mathbf{s}_1,\mathbf{o}_2) \\
~\mathcal{DP}(\mathbf{s}_2,\mathbf{o}_2) \\
\vdots \\
~\mathcal{DP}(\mathbf{s}_n,\mathbf{o}_2)
\end{bmatrix} \,. $$
Hence, the maximum rank of $M$ is $o_2$, since $\mathbf{o}_2\in O_2$. This observation provides attacker a min-rank instance to find ${o_{2}}_{i}$'s in $\mathbb{F}_q$.

To enhance the performance of the simple attack, Beullens combined the rectangular min-rank attack with the simple attack  \cite{beullens2022breaking}. Like earlier, the attacker fixes $\mathbf{v}$ to get a linear map $\mathcal{DP}_\mathbf{v}$. This helps to find $\mathbf{o}_2\in O_2$ using $\mathcal{DP}_\mathbf{v}(\mathbf{o}_2)=0$.

This system of linear equations reduces the number of matrices by $m$ in the rectangular min-rank instance. Thus, the basis of Ker($\mathcal{DP}_{\mathbf{v}})$ is $\mathbf{b}_1,\cdots,\mathbf{b}_{n-m}$. Hence, the new min-rank instance has $n-m$ matrices $\widetilde{M}_i$, where 
$$\widetilde{{M}_i} \,:=\, \sum_{j=1}^nb_{ij}M_j \,:=\,
 \begin{bmatrix}
~\mathcal{DP}(\mathbf{s}_1,\mathbf{b}_i) \\
~\mathcal{DP}(\mathbf{s}_2,\mathbf{b}_i) \\
\vdots \\
~\mathcal{DP}(\mathbf{s}_n,\mathbf{b}_i)
\end{bmatrix} , \quad \mbox{ for }i=1 \mbox{ to } n-m \,.$$
If $\mathbf{y}$ is a solution of the new min-rank problem having $n-m$ matrices then $\mathbf{o}_2=\sum_{i=1}^{n-m}y_i \mathbf{b}_i$ is a solution of the old min-rank problem. 
Hence, the attack needs to be repeated approximately $q$ times, until it finds $\mathbf{o}_2\;\in\mbox{ Ker}{(\mathcal{DP}_\mathbf x)}\;\cap \;O_2\neq\;\{0\}$.

 \medskip{\textbf{Attack Complexity.}} 
   The number of field multiplications required to perform this attack is 
 $$3\cdot q\cdot (n-m-1)(o_2+1)\binom{n}{r}^2\cdot\binom{n-m+b-3}{b}^3$$ where $b$ is the operating degree for the algorithm \cite{bardet2020improvements}. 

 %\smallskip \noindent\textbf{\em Rectangular min-rank attack against VDOO.} This combined attack can be applied to our proposal also. Like simple attack, attacker expected a good guess for $\mathcal{DP}_\textbf{x}$. Then using this information, the attacker is able to get min-rank instances of 60 matrices. Each matrix has $n-1$ rows and $m$-columns and the span of these matrices has a matrix of rank $o_2$. Once a vector $\textbf{o}\in O_2$ is found, then it will be easy for an attacker to recover all vectors (as described in the simple attack against VDOO).
 
 \textit{\textbf {Example for SL-1 parameters.}} The attacker needs to guess a good $\mathcal{DP}_\textbf{x}$. After then the attacker gets a min-rank instance of 60 matrices which has 159 rows and 100 columns and the span of these matrices has a matrix of rank 36. Bardet et al.'s~ \cite{bardet2020improvements} algorithm provides an efficient way to solve this min-rank instance. This computation needs $2^{133}$-field operations.  
\subsection{Kipnis-Shamir Attack on VDOO}
The attacker targeting VDOO can employ a technique similar to the one devised by Kipnis and Shamir~\cite{kipnis1998cryptanalysis} to retrieve the subspace $O_2$. This approach effectively aids in the separation of oil and vinegar variables, ultimately leading to the recovery of the private key. The complexity of this attack can be roughly estimated as $O(o_2^4 \cdot q^{n - o_2 - 1})$ field multiplications. To expedite this assault, the attacker leverages Grover's algorithm, which serves to reduce the complexity to $O(o_2^4 \cdot q^{(n - o_2 - 1)/2})$.

{\bf \textit{Example for SL-1 parameters.} }Attacker needs to perform approximately $2^{348}$-field operations in classical settings and $2^{174}$-field operations in quantum computer.
 \subsection{Intersection Attack on VDOO}
Beullens introduced the intersection attack~\cite{beullens2021improved}, which effectively reduced the claimed security level of the Rainbow signature scheme by approximately 20 bits compared to the original design. In this attack, Beullens improved upon the Rainbow band separation attack~\cite{ding2008new} using the analysis proposed by Perlner~\cite{perlner2020rainbow}. The intersection attack helps to identify $k$-vectors simultaneously within the oil-space $O_2$ by solving a system of quadratic equations for a vector within the intersection $\cap_{i=1}^kL_iO_2$, where $L_i$'s are invertible matrices. This attack performs well when the intersection is non-empty, which occurs when $n < \frac{2k-1}{k-1}o_2$. The computational cost of this attack involves solving a quadratic system with $\binom{k+1}{2}^2o_2-2\binom{k}{2}$ equations in $k(no_2 )-(2k-1)o_2$ variables.

However, in the case of VDOO where $n \geq 3o_2$, there is no guarantee that the subspace (for more details, see~\cite{beullens2021improved}) namely $L_iO_2\,\cap\, L_jO_2$ will exist. Consequently, the attack becomes probabilistic for VDOO and will succeed with a probability of $\frac{1}{q^{(n-3o_2+1)}}$.

{\bf \textit{Example for SL-1 parameters.} }The complexity to break SL-1 parameters, attacker needs $2^{131}$-field multiplications.
%to attack the SL-1 parameter set.
%\vspace{-1em}
\subsection{Quantum Attacks}
The attacker can accelerate certain aspects of the classical attacks using a quantum computer. For MQ- or OV-based schemes the only quantum algorithm that can help in cryptanalysis is Grover's search~\cite{grover1996fast}. This algorithm reduces the search space, thereby reducing the number of field multiplications by a factor of $q^{k/2}$. This specifically does not threaten the post-quantum security of our scheme~\cite{chen2017nist}.
\subsection{Provable security: \textsc{EUF-CMA} Security}
Our VDOO scheme, similar to UOV, Rainbow, and other UOV-based signature schemes, offers universal unforgeability \cite{ding2005rainbow}. Like these other schemes, we incorporate a salt in the signature generation process to demonstrate the {\sc EUF-CMA} security of our scheme. We have followed the established methodology for this purpose, as seen in prior work such as \cite{sakumoto2011provable,beullens2022mayo}. %While we haven't provided a full proof here, we have outlined the proof. 
Here, we have only provided an outline of the proof. The full proof can be done using similar strategies as Mayo~\cite{beullens2022mayo}, QR-UOV~\cite{furue2023qr}, PROV~\cite{faugereprov}, etc. 
Our security proof relies on the well-understood hardness of the UOV problem. We begin by defining the UOV problem and then introduce the VDOO problem.

For security reasons, we recommend that each salt value should be used for no more than one signature. Consequently, we fix the salt length at 16 bytes, assuming up to $2^{64}$ signature generations within the system~\cite{chen2017nist}.
%Since VDOO could be depicted as a multi-layer Rainbow utilizing UOV layers, we can make VDOO \textsc{EUF-CMA} secure with a change like \cite{sakumoto2011provable}. \cite{sakumoto2011provable} has detailed  the proof for UOV. The same technique can be used to assert that the \textsc{EUF-CMA} security  of the VDOO is the same as  the security of the original Rainbow. 

%In this scenario adversary gets a VDOO public key and access to the corresponding signing oracle that can be required at most $2^{64}$ times. Goal is to generate a valid signature for a new message \cite{chen2017nist}. 
\begin{definition}[UOV Problem] Suppose $\mathsf{UOV}_{(n,v,o,q)}$ denotes a family of UOV public polynomial maps where $n$ is the number variables, $v+o$ is number of equations and $q$ is the size of the finite field, and $\mathsf{MQ}_{(q,n,m)}$ denotes a family of random quadratic systems with $n$ unknowns and $m$ constraints over $\fq$. The UOV problem asks to distinguish $\p$ from 
$\mathsf{UOV}_{(q,n,v,o,)}$ and $\mathsf{MQ}_{(q,n,m)}$.
Suppose $\mathcal{A}_{\mathsf{UOV}}$ be the adversary solves the distinguishing problem and it has a distinguishing advantage as:
$$\mathsf{Adv}_{\mathsf{UOV}}(\mathcal{A}_{{\mathsf{UOV}}}
)=\big|\mbox{Pr}[\mathcal{A}_{\mathsf{UOV}}(\p) =1~|~\p\in \mathsf{MQ}]-\mbox{Pr}[\mathcal{A}_{\mathsf{UOV}}(\p) =1~|~\p\in \mathsf{UOV}]\big|$$\end{definition}
It is widely believed that there is no probabilistic polynomial-time adversary, including quantum adversaries, denoted as $\mathcal{A}$, that can efficiently solve the UOV problem.
\begin{definition}[\textbf{VDOO Problem}] Suppose $\mathsf{VDOO}$ be a family of VDOO public polynomial map. Now given a random $\p\in\mathsf{VDOO}\mbox{ and }\mathbf{t}\in\fqm$ VDOO problem asks to find $\mathbf{s}$ such that $\p(\mathbf{s})=\mathbf{t}$. If $\mathcal{A}$ is such an adversary to compute the inverse of the VDOO public map then the advantage of this computation is 
$$\mathsf{Adv}_{\mathsf{VDOO}}(\mathcal{A}_ \mathsf{VDOO})=\mbox{Pr}\;[\;\p(\mathbf{{s}})=\mathbf{t}~|~ \p\in\mathsf{VDOO}, ~\mathcal{A}_\mathsf{VDOO}(\p,\mathbf{t})=\mathbf{s}\;]$$\end{definition}

%\begin{definition}[\textbf{\textsc{EUF-CMA} Security.}] We assume $\mathcal{H}$ be the random oracle and $\mathcal{A}$ is the adversary. The {\it advantage }of \textsc{EUF-CMA} security of signature scheme VDOO=\{{\sf VDOOKeyGen, VDOOSign, VDOOVerif}\} is defined as $$ \mathsf{Adv}_{{\sf VDOO}}^{\textsc{\tiny EUF-CMA}} =\mbox{Pr}[{\sf VDOOVerif(pk,msg,sig)}=\perp \& {\sf VDOOSign}^{\mathcal{H}}(sk,\cdot) \mbox{ was never queried on message }]$$ where $(pk,s k)\leftarrow {\sf VDOOKeyGen()}$ and $(msg,~sig)\leftarrow \mathcal{A}^{\mathcal{H}, {\sf VDOOSign (sk,\cdot)}}(pk)$\end{definition}

%\textbf{Modified Signature Generation} To achieve \textsc{EUF-CMA} security, we slightly change in the the signature generation. As a consequence of this change, our verification algorithm will change and signature size also increases. 

Now we are going to state our main theorem which establishes the {\sc EUF-CMA} security of the  VDOO. To understand the security notion, we refer to ~\cite{beullens2022mayo,sakumoto2011provable,kosuge2022probabilistic}.
 \begin{theorem}
 Suppose the adversary $\mathcal{A}$ runs in time $T$ to solve the {\sc EUF-CMA} game of VDOO in the random oracle model. This adversary makes $q_s$ signing queries and $q_h$ random oracle queries. Then there exists $\mathcal{A}_{UOV}$ and $\mathcal{A}_{VDOO}$ running in time $T+O((q_s+q_t)\cdot \;poly(q,v,d,o_1,o_2))$ with 
 \begin{align*}\mathsf{Adv}^{\mbox{{\sc\tiny EUF-CMA}}}_{\mathsf{VDOO}} (\mathcal{A}) \;\leq\; &\;\mathsf{Adv}_{\mathsf{UOV}_{(q,v',o')}} (\mathcal{A}_{UOV}) \;+\; q_h\cdot  \mathsf{Adv}_{\mathsf{VDOO}_{(q,v,d,o_1,o_2)}} (\mathcal{A}_{VDOO})\\ & \;+\;(q_s+q_h)q_s\cdot 2^{-|salt|} \;+\; q^{-m}\,.\end{align*}\end{theorem}
\emph{Proof idea.} Here, we informally sketch the proof. We can adopt the proof methodology used in Mayo (see theorem $6$ from~\cite{beullens2022mayo}). In the first step, we can establish a reduction from the {\sc EUF-CMA} security of the VDOO signature scheme to {\sc EUF-KOA} (Existential unforgeability against key-only attack) security by simulating the signing oracle. Note that, the adversary does not have access to the signing oracle in the {\sc EUF-KOA} game. Once this reduction is established, we can easily show a reduction from the UOV problem and VDOO problem to the {\sc EUF-KOA} security game in the second step. Like the security proof of Mayo~\cite{beullens2022mayo}, we can use the hybrid proof system to establish both reductions. This proof style has also been adopted by many state-of-the-art OV-based constructions \cite{furue2023qr,furue2021new,faugereprov,ding2023tuov}. Finally, we can combine both of these two steps to establish the above theorem.
\section{Parameters and Performance}\label{sec:parms}
%We select parameters based on the most effective known attacks against the proposed scheme. Initially, we set these parameters and then proceed to compare our scheme with the current state-of-the-art.
This section describes our chosen parameters based on the security analysis described in Section.~\ref{sec:sec_analysis_params}. We assess the \emph{practicality} of the VDOO signature scheme, which involves a finely tuned trade-off among computation time, security, and communication costs. 
%Practicality in this context refers to the trade-off between the time it takes to perform computations and the cost of communication. 
For most multivariate schemes, computation time is dominated by either the Gaussian elimination (solving linear system \footnote{$\mathsf{GE}_{(q,n)}$: Gaussian elimination on a linear system with $n$ unknowns and $n$ linear equation over $\fq$. This computation needs $O(n^3)$-field operations.}) or the Gr\"{o}bner basis method (solving quadratic system \footnote{$\mathsf{XL}_{(q,n)}$: eXtended Linearization or Gr\"{o}bner basis method to solve a quadratic system of $n$ variables and $n$ constraints over $\fq$. This computation needs $2^{2^n}$-field operations.}). Communication cost is proportional to signature size $+$ public key size.
%We also compare our scheme with state-of-the-art multivariate and other post-quantum signature schemes.%\vspace{-1em}
%\vspace{-2pt}
\subsection{Parameter Selection}
%In Table~\ref{tab:vdoo}, we calculate the key and signature sizes for VDOO across a wide range of parameter values. These parameters are chosen following NIST guidelines \cite{chen2017nist}, where the ``parameter column'' is defined as $\mathsf{params}=(q,v,d,o_1,o_2)$, as explained earlier. 
Table.~\ref{tab:vdoo}, shows the signature, public-key, and private-key sizes of VDOO for different security levels as determined by the parameter tuple $(q,v,d,o_1,o_2)$. We follow the NIST classification~\cite{chen2017nist} to categorize the parameters. 
%The ``attack complexity'' column presents the complexity of two primary attacks: the simple attack \cite{beullens2022breaking} (SA) and the rectangular min-rank attack \cite{beullens2021improved} (RA), represented as a tuple. Since both of these attacks exhibit the best complexity compared to others, we provide the complexity in terms of the number of field multiplications required for their execution.
We consider the complexity of two primary attacks: the simple attack~\cite{beullens2022breaking} (SA) and the rectangular min-rank attack~\cite{beullens2021improved} (RA). From the attacker's point of view, these two attacks exhibit the most optimistic complexity among all other known attacks. Here, the complexity represents the number of field multiplications required for their execution.
\begin{table}[]
\centering
\begin{tabular}{@{}c|c|c|c|c|c@{}}
\toprule
 \begin{tabular}[c]{@{}c@{}}Security\\ level (B)\end{tabular} &
  \begin{tabular}[c]{@{}c@{}}$\mathsf{params}$ \\  $(q,\;v,\;d,\;o_1,\;o_2)$\end{tabular} &
  \begin{tabular}[c]{@{}c@{}}\textbf{Signature}\\ \textbf{size (B)}\end{tabular} &
  \begin{tabular}[c]{@{}c@{}}\textbf{Private key} \\ \textbf{size (KB)}\end{tabular} &
  \begin{tabular}[c]{@{}c@{}}\textbf{Public key}\\ \textbf{size (KB)}\end{tabular} &
  \begin{tabular}[c]{@{}c@{}}\textbf{Attacks}\\ \textbf{(SA, RA)}\end{tabular} \\ \midrule
SL-I & $~(16,\;60,\;30\;34,\;36)~~$   & 96  & 243  & 236  & $~~(134,\,138)~~$\\ \midrule
SL-III & $~(256,\;100,\;30,\;40,\;40)~~$ & 226 & 1056 & 2437 & $~~(207,\, 191)~~$ \\ \midrule
SL-V & $~(256,\;120,\;50,\;60,\;70)~~$ & 316 & 3524 & 8127 & $~~(270,\;264)~~$  \\ \bottomrule
\end{tabular}
\medskip\caption{VDOO parameter set for different NIST prescribed security level}
\label{tab:vdoo}
\end{table}
\vspace{-3pt}
\subsection{Comparison with other post-quantum schemes}%\vspace{-0.1mm}
%In this section, we assess the "practicality" of the VDOO signature scheme, which involves balancing computation time and communication costs. Practicality in this context refers to the trade-off between the time it takes to perform computations and the cost of communication. For most multivariate schemes, computation time typically involves Gaussian elimination (solving linear system \footnote{$\mathsf{GE}_{(q,n)}$: Gaussian elimination on a linear system with $n$ unknowns and $n$ linear equation over $\fq$. This computation needs $O(n^3)$-field operations.}) or the Gr\"{o}bner basis method (solving quadratic system \footnote{$\mathsf{XL}_{(q,n)}$: eXtended Linearization or Gr\"{o}bner basis method to solve a quadratic system of $n$ variables and $n$ constraints over $\fq$. This computation needs $2^{2^n}$-field operations.}). Communication cost encompasses both signature size and public key size.

%In the field of post-quantum cryptography, numerous signature schemes were submitted to the NIST post-quantum standardization project, with eleven of them originating from multivariate cryptography. 
In response to the NIST's last~\cite{chen2017nist} and the latest~\cite{nistadditional} standardization call multiple post-quantum signatures schemes have been proposed based on MQ problem or its derivatives. For our comparative analysis, we focus on schemes with small signature sizes and well-established hardness assumptions only in Table.~\ref{tab:compmul}. For fairness, we compare with the parameters which provide at least 128-bit of classical security~\cite{chen2017nist}. For details about the parameters of a scheme and their role in security and key sizes we kindly request interested readers to the original publications. 
%To better understand the parameter representations of these other schemes, we refer to their respective references (also see~\cite{nistadditional}).

%In this section, we compare our proposal with Rainbow (with updated parameters \cite{nistgoogle,ding2005rainbow}), IPRainbow  \cite{cartor2022iprainbow}, UOV, Mayo \cite{beullens2022mayo} (see Table \ref{tab:compmul}). 

% Please add the following required packages to your document preamble:
% \usepackage{longtable}
% Note: It may be necessary to compile the document several times to get a multi-page table to line up properly

% Please add the following required packages to your document preamble:
% \usepackage{longtable}
% Note: It may be necessary to compile the document several times to get a multi-page table to line up properly
\begin{longtable}[c]{|c|c|c|c|}
\hline
\begin{tabular}[c]{@{}c@{}}{\bf Signature} \\ {\bf schemes}\end{tabular} &
  \begin{tabular}[c]{@{}c@{}}{\bf Computational}\\ {\bf bottleneck}\end{tabular} &
  \begin{tabular}[c]{@{}c@{}}{\bf Signature}\\{\bf size (B)}\end{tabular} &
  \begin{tabular}[c]{@{}c@{}}{\bf Public key}\\ {\bf size (KB)}\end{tabular} \\ \hline
\endfirsthead
\multicolumn{4}{c}%
{{\bfseries Table \thetable\ continued from previous page}} \\
\endhead
\begin{tabular}[c]{@{}c@{}}VDOO\\ $(16,40,30,34,36)$\end{tabular}   & $\mathsf{GE}_{(16,34)}$, $\mathsf{GE}_{(16,36)}$      & 96  & 238    \\ \hline
\begin{tabular}[c]{@{}c@{}}Rainbow \cite{ding2005rainbow,nistgoogle}\\ $(256,148,80,48)$\end{tabular} & $\mathsf{GE}_{(256,32)}$,  $\mathsf{GE}_{(256,48)}$ & 164 & 258    \\ \hline
\begin{tabular}[c]{@{}c@{}}IPRainbow \cite{cartor2022iprainbow}\\ $(257,32,32,38,7)$\end{tabular} &
  \begin{tabular}[c]{@{}c@{}}$\mathsf{GE}_{(257,32)}$, $\mathsf{GE}_{(257,38)}$, \\ $\mathsf{XL}_{(257,7)}$\end{tabular} &
  120 &
  342.784 \\ \hline
\begin{tabular}[c]{@{}c@{}}Mayo \cite{beullens2022mayo}\\ $(16, 66,65,7,11)$\end{tabular}   & $\mathsf{GE}_{(16,65)}$                             & 387 & 1      \\ \hline
\begin{tabular}[c]{@{}c@{}}QR-UOV \cite{furue2023qr,furue2021new}\\ $(7,740,100,10)$\end{tabular}   & $\mathsf{GE}_{(7,100)}$                             & 331 & 20.657 \\ \hline
\begin{tabular}[c]{@{}c@{}}PROV\cite{faugereprov}\\ $(136, 46,8)$\end{tabular}        & $\mathsf{GE}_{(8,46)}$                              & 160 & 68.326 \\ \hline
\begin{tabular}[c]{@{}c@{}}TUOV \cite{ding2023tuov}\\ $(16, 160, 64, 32)$\end{tabular}   & $\mathsf{GE}_{(16,64)}$ +$\mathsf{GE}_{(16,32)}$                            & 80  & 65.552 \\ \hline
\begin{tabular}[c]{@{}c@{}}VOX \cite{voxfranceprincipal}\\ $(251,8,9,6,6)$\end{tabular}       & $\mathsf{XL}_{(251,6)}$                             & 102 & 9.1    \\ \hline
\begin{tabular}[c]{@{}c@{}}UOV \cite{beullensuov}\\ $(256,160,64,16)$\end{tabular}     & $\mathsf{GE}_{(256,64)}$                            & 96  & 66.576 \\ \hline
\caption{Compare with other multivariate signature for security level one (at least 128-bit) \cite{chen2017nist}}
\label{tab:compmul}\\
\end{longtable}

In Table.~\ref{tab:compsign}, we compare VDOO with recently standardized Crystals Dilithium~\cite{dilithium}, Falcon~\cite{web:falcon}, SPHINICS+ \cite{web:sphincs} and recently submitted some signature schemes (see \cite{nistadditional}) which are not based on MQ problem.

% Please add the following required packages to your document preamble:
% \usepackage{longtable}
% Note: It may be necessary to compile the document several times to get a multi-page table to line up properly
\begin{longtable}[c]{|c|c|c|c|c|c|c|}
\hline
\begin{tabular}[c]{@{}c@{}}{\bf Comparisons/}\\{\bf  Algorithms}\end{tabular} & VDOO    & \begin{tabular}[c]{@{}c@{}}Crystals \\ Dilithium\end{tabular} & Falcon     & Sphincs+ & FuLeeca & LESS  \\ \hline
\endfirsthead
\multicolumn{7}{c}%
{{\bfseries Table \thetable\ continued from previous page}} \\
\endhead
\begin{tabular}[c]{@{}c@{}}Signature \\ size (B)\end{tabular}     & 96      & 2420                                                          & 666        & 7856     & 1100    & 8400  \\ \hline
\begin{tabular}[c]{@{}c@{}}Public key \\ size (B)\end{tabular}    & 23813   & 1312                                                          & 897        & 32       & 1318    & 13700 \\ \hline
\begin{tabular}[c]{@{}c@{}}{\bf Comparisons}/\\{\bf Algorithms}\end{tabular} & SQISign & Hawk                                                          & ASCON-Sign & MIRA     & MiRitH  & RYDE  \\ \hline
\begin{tabular}[c]{@{}c@{}} Signature\\  size (B)\end{tabular}     & 177     & 555                                                           & 7856       & 7376     & 7661    & 7446  \\ \hline
\begin{tabular}[c]{@{}c@{}} Public key \\ size (B)\end{tabular}    & 64      & 1024                                                          & 32         & 84       & 129     & 86    \\ \hline
\caption{Comparisons with other signatures for NIST security level 1}
\label{tab:compsign}\\
\end{longtable}

From the above tables, it is evident that VDOO outperforms the majority of existing multivariate signature schemes. This superiority stems from the smaller number of variables involved in Gaussian eliminations in VDOO. Furthermore, the signature generation process in VDOO does not rely on the Gr\"{o}bner basis technique, which further confirms its practicality. Further Table.~\ref{tab:compsign} illustrates that VDOO has one of the smallest signature sizes with respect to other quantum-safe signature schemes.

%the latter needs Gr\"{o}bner basis algorithm in the signature phase. Gr\"{o}bner basis algorithm is one of the most expensive algorithms; due to this heavy computation IPRainbow is approximately 50 times slower than our scheme. Further, our diagonal trick reduces the size of Gaussian elimination in Rainbow. Hence, VDOO performs at least as well as both Rainbow and IPRainbow signature schemes. Mayo is one of the beautiful signature schemes designed in 2021. But it heavily suffers from large signature size. Considering all aspects, VDOO is a practical signature scheme.  
%\vspace{-12pt}
\section{Conclusion}\label{sec:conclusion}
%\vspace{-8pt}
We have introduced a post-quantum signature algorithm, leveraging well established cryptanalysis techniques to devise a parameter set for VDOO. In order to ensure a minimum of 128-bit security, our scheme achieves a compact 96-byte signature size, which outperforms numerous existing signature schemes. Nonetheless, it does grapple with a sizable public key size, a challenge that is prevalent in a significant number of multivariate signature schemes.

Our immediate future endeavors will be centered around further compressing the public key size within the VDOO scheme. Additionally, we intend to delve into the exploration of VDOO's security within the quantum random oracle model (QROM). Subsequently, our focus will shift towards realizing hardware implementations and assessing potential physical attacks against our scheme.
\section{Acknowledgements}\label{sec:ackn}
Authors thanks to anonymous reviewers for their valuable feedback. AG wish thanks the Tata Consultancy Service for funding. N.S. thanks the funding support from DST-SERB (CRG/2020/000045) and N.Rama Rao Chair (CSE-IITK).
%
% ---- Bibliography ----
%
% BibTeX users should specify bibliography style 'splncs04'.
% References will then be sorted and formatted in the correct style.
%

 \bibliographystyle{splncs04}
 \bibliography{main}

\begin{thebibliography}{10}
\providecommand{\url}[1]{\texttt{#1}}
\providecommand{\urlprefix}{URL }
\providecommand{\doi}[1]{https://doi.org/#1}

\bibitem{agrawal2005automorphisms}
Agrawal, M., Saxena, N.: Automorphisms of finite rings and applications to
  complexity of problems. In: Annual Symposium on Theoretical Aspects of
  Computer Science. pp. 1--17. Springer (2005)

\bibitem{agrawal2006equivalence}
Agrawal, M., Saxena, N.: Equivalence of $\mathbb{F}$-algebras and cubic forms.
  In: Annual Symposium on Theoretical Aspects of Computer Science. pp.
  115--126. Springer (2006)

\bibitem{nist_final_report}
Alagic, G., Apon, D., Cooper, D., Dang, Q., Dang, T., Kelsey, J., Lichtinger,
  J., Liu, Y.K., Miller, C., Moody, D., Peralta, R., Perlner, R., Robinson, A.,
  Smith-Tone, D.: {Status Report on the Third Round of the NIST Post-Quantum
  Cryptography Standardization Process}. Online. Accessed 26th June, 2023
  (2022), \url{https://nvlpubs.nist.gov/nistpubs/ir/2022/NIST.IR.8413-upd1.pdf}

\bibitem{web:sphincs}
Aumasson, J.P., Bernstein, D.J., Beullens, W., Dobraunig, C., Eichlseder, M.,
  Fluhrer, S., Gazdag, S.L., H\:ulsing, A., Kampanakis, P., Kölbl, S., Lange,
  T., Martin M.~Lauridsen, F.M., Niederhagen, R., Rechberger, C., Rijneveld,
  J., Schwabe, P., Westerbaan, B.: Sphincs+ submission to the nist post-quantum
  project, v.3.1 (2018),
  \url{https://sphincs.org/data/sphincs+-r3.1-specification.pdf}, [Online;
  accessed 10-June-2023]

\bibitem{baena2022improving}
Baena, J., Briaud, P., Cabarcas, D., Perlner, R., Smith-Tone, D., Verbel, J.:
  Improving support-minors rank attacks: Applications to {GeMSS and Rainbow}.
  In: Annual International Cryptology Conference. pp. 376--405. Springer (2022)

\bibitem{bardet2002algebraic}
Bardet, M., Bros, M., Cabarcas, D., Gaborit, P., Perlner, R., Smith-Tone, D.,
  Tillich, J.P., Verbel, J.: Algebraic attacks for solving the rank decoding
  and min-rank problems without {G}r{\"o}bner basis (2020). Preprint available
  on https://arxiv. org/pdf/2002.08322. pdf  \textbf{3},  22--30

\bibitem{bardet2020improvements}
Bardet, M., Bros, M., Cabarcas, D., Gaborit, P., Perlner, R., Smith-Tone, D.,
  Tillich, J.P., Verbel, J.: Improvements of algebraic attacks for solving the
  rank decoding and {MinRank} problems. In: International Conference on the
  Theory and Application of Cryptology and Information Security. pp. 507--536.
  Springer (2020)

\bibitem{bernstein2017classic}
Bernstein, D.J., Chou, T., Lange, T., von Maurich, I., Misoczki, R.,
  Niederhagen, R., Persichetti, E., Peters, C., Schwabe, P., Sendrier, N.,
  et~al.: {Classic McEliece: Conservative Code-based Cryptography}. NIST
  submissions  (2017)

\bibitem{bettale2009hybrid}
Bettale, L., Faugere, J.C., Perret, L.: Hybrid approach for solving
  multivariate systems over finite fields. Journal of Mathematical Cryptology
  \textbf{3}(3),  177--197 (2009)

\bibitem{beullens2021improved}
Beullens, W.: Improved cryptanalysis of {UOV} and {R}ainbow. In: Annual
  International Conference on the Theory and Applications of Cryptographic
  Techniques. pp. 348--373. Springer (2021)

\bibitem{beullens2022breaking}
Beullens, W.: Breaking {R}ainbow takes a weekend on a laptop. Cryptology ePrint
  Archive  (2022)

\bibitem{beullens2022mayo}
Beullens, W.: Mayo: practical post-quantum signatures from oil-and-vinegar
  maps. In: Selected Areas in Cryptography: 28th International Conference,
  Virtual Event, September 29--October 1, 2021, Revised Selected Papers. pp.
  355--376. Springer (2022)

\bibitem{beullensuov}
Beullens, W., Chen, M.S., Ding, J., Gong, B., Kannwischer, M.J., Patarin, J.,
  Peng, B.Y., Schmidt, D., Shih, C.J., Tao, C., Yang, B.Y.: {UOV: {U}nbalanced
  {O}il and {V}inegar Algorithm Specifications and Supporting Documentation
  Version 1.0} (2018),
  \url{https://csrc.nist.gov/csrc/media/Projects/pqc-dig-sig/documents/round-1/spec-files/UOV-spec-web.pdf},
  [Online; accessed 5-September-2023]

\bibitem{billet2006cryptanalysis}
Billet, O., Gilbert, H.: Cryptanalysis of {R}ainbow. In: International
  Conference on Security and Cryptography for Networks. pp. 336--347. Springer
  (2006)

\bibitem{Kyber-Kem}
Bos, J., Ducas, L., Kiltz, E., Lepoint, T., Lyubashevsky, V., Schanck, J.M.,
  Schwabe, P., Seiler, G., Stehlé, D.: {CRYSTALS -- Kyber: a CCA-secure
  module-lattice-based KEM}. Cryptology ePrint Archive, Report 2017/634 (2017),
  \url{https://ia.cr/2017/634}

\bibitem{buss1999computational}
Buss, J.F., Frandsen, G.S., Shallit, J.O.: The computational complexity of some
  problems of linear algebra. Journal of Computer and System Sciences
  \textbf{58}(3),  572--596 (1999)

\bibitem{cartor2022iprainbow}
Cartor, R., Cartor, M., Lewis, M., Smith-Tone, D.: {IPR}ainbow. In:
  International Conference on Post-Quantum Cryptography. pp. 170--184. Springer
  (2022)

\bibitem{decru_sidh}
Castryck, W., Decru, T.: An efficient key recovery attack on {SIDH}. In: Hazay,
  C., Stam, M. (eds.) Advances in Cryptology - {EUROCRYPT} 2023 - 42nd Annual
  International Conference on the Theory and Applications of Cryptographic
  Techniques, Lyon, France, April 23-27, 2023, Proceedings, Part {V}. Lecture
  Notes in Computer Science, vol. 14008, pp. 423--447. Springer (2023).
  \doi{10.1007/978-3-031-30589-4\_15},
  \url{https://doi.org/10.1007/978-3-031-30589-4\_15}

\bibitem{chen2017nist}
Chen, L., Moody, D., Liu, Y.: {NIST} post-quantum cryptography standardization.
  Transition  \textbf{800}, ~131A (2017)

\bibitem{nistadditional}
Chen, L., Moody, D., Liu, Y.K.: Post-quantum cryptography: Digital signature
  schemes. round 1 additional signatures,
  \url{https://csrc.nist.gov/Projects/pqc-dig-sig/round-1-additional-signatures}

\bibitem{courtois2000efficient}
Courtois, N., Klimov, A., Patarin, J., Shamir, A.: Efficient algorithms for
  solving overdefined systems of multivariate polynomial equations. In:
  International Conference on the Theory and Applications of Cryptographic
  Techniques. pp. 392--407. Springer (2000)

\bibitem{de2020sqisign}
De~Feo, L., Kohel, D., Leroux, A., Petit, C., Wesolowski, B.: {SQIS}ign:
  {C}ompact {P}ost-quantum {S}ignatures from {Q}uaternions and {I}sogenies. In:
  Advances in Cryptology--ASIACRYPT 2020: 26th International Conference on the
  Theory and Application of Cryptology and Information Security, Daejeon, South
  Korea, December 7--11, 2020, Proceedings, Part I 26. pp. 64--93. Springer
  (2020)

\bibitem{ding2023tuov}
Ding, J.: Tuov: Triangular unbalanced oil and vinegar  (2023)

\bibitem{ding2005rainbow}
Ding, J., Schmidt, D.: Rainbow, a new multivariable polynomial signature
  scheme. In: International conference on applied cryptography and network
  security. pp. 164--175. Springer (2005)

\bibitem{ding2008new}
Ding, J., Yang, B.Y., Chen, C.H.O., Chen, M.S., Cheng, C.M.: New
  differential-algebraic attacks and reparametrization of {R}ainbow. In:
  International Conference on Applied Cryptography and Network Security. pp.
  242--257. Springer (2008)

\bibitem{dilithium}
Ducas, L., Kiltz, E., Lepoint, T., Lyubashevsky, V., Schwabe, P., Seiler, G.,
  Stehlé, D.: Crystals-dilithium: A lattice-based digital signature scheme.
  IACR Transactions on Cryptographic Hardware and Embedded Systems
  \textbf{2018}(1),  238–268 (Feb 2018).
  \doi{10.13154/tches.v2018.i1.238-268},
  \url{https://tches.iacr.org/index.php/TCHES/article/view/839}

\bibitem{faugere1999new}
Faugere, J.C.: A new efficient algorithm for computing {G}r{\"o}bner bases
  ({F4}). Journal of pure and applied algebra  \textbf{139}(1-3),  61--88
  (1999)

\bibitem{faugere2002new}
Faugere, J.C.: A new efficient algorithm for computing {G}r{\"o}bner bases
  without reduction to zero {(F5)}. In: Proceedings of the 2002 {I}nternational
  {S}ymposium on {Symbolic and Algebraic Computation}. pp. 75--83 (2002)

\bibitem{faugereprov}
Faugere, J.C., Fouque, P.A., Macario-Rat, G., Minaud, B., Patarin, J.: {PROV:
  PR}ovable unbalanced {O}il and {V}inegar specification v1. 0--06/01/2023

\bibitem{faugere2008cryptanalysis}
Faugere, J.C., Levy-dit Vehel, F., Perret, L.: Cryptanalysis of {M}in-{R}ank.
  In: Annual International Cryptology Conference. pp. 280--296. Springer (2008)

\bibitem{sidh}
Feo, L.D., Jao, D., Pl{\^{u}}t, J.: Towards quantum-resistant cryptosystems
  from supersingular elliptic curve isogenies. J. Math. Cryptol.
  \textbf{8}(3),  209--247 (2014). \doi{10.1515/jmc-2012-0015},
  \url{https://doi.org/10.1515/jmc-2012-0015}

\bibitem{web:falcon}
Fouque, P.A., Hoffstein, J., Kirchner, P., Lyubashevsky, V., Pornin, T., Prest,
  T., Ricosset, T., Seiler, G., Whyte, W., Zhang, Z.: Falcon: Fast-fourier
  lattice-based compact signatures over ntru (2018),
  \url{https://falcon-sign.info/}, [Online; accessed 10-June-2023]

\bibitem{voxfranceprincipal}
France, T.D., Faug{\`e}re, J.C., Fouque, P.A., Goubin, L., Larrieu, R.,
  Macario-Rat, G., Minaud, B.: Principal submitter: Jacques patarin

\bibitem{furue2023qr}
Furue, H., Ikematsu, Y., Hoshino, F., Kiyomura, Y., Saito, T., Takagi, T.:
  Qr-uov  (2023)

\bibitem{furue2021new}
Furue, H., Ikematsu, Y., Kiyomura, Y., Takagi, T.: A new variant of unbalanced
  oil and vinegar using quotient ring: {QR-UOV}. In: Advances in
  Cryptology--ASIACRYPT 2021: 27th International Conference on the Theory and
  Application of Cryptology and Information Security, Singapore, December
  6--10, 2021, Proceedings, Part IV 27. pp. 187--217. Springer (2021)

\bibitem{nistgoogle}
Groups, G.: Rainbow round3 official comment (2022)

\bibitem{grover1996fast}
Grover, L.K.: A fast quantum mechanical algorithm for database search. In:
  Proceedings of the twenty-eighth annual {ACM Symposium on Theory of
  Computing}. pp. 212--219 (1996)

\bibitem{johnson1979computers}
Johnson, D.S., Garey, M.R.: Computers and Intractability: A Guide to the Theory
  of NP-completeness. WH Freeman (1979)

\bibitem{kipnis1999unbalanced}
Kipnis, A., Patarin, J., Goubin, L.: Unbalanced oil and vinegar signature
  schemes. In: International Conference on the Theory and Applications of
  Cryptographic Techniques. pp. 206--222. Springer (1999)

\bibitem{kipnis1998cryptanalysis}
Kipnis, A., Shamir, A.: Cryptanalysis of the {O}il and {V}inegar signature
  scheme. In: Annual international cryptology conference. pp. 257--266.
  Springer (1998)

\bibitem{kipnis1999cryptanalysis}
Kipnis, A., Shamir, A.: Cryptanalysis of the {HFE} public key cryptosystem by
  relinearization. In: Annual International Cryptology Conference. pp. 19--30.
  Springer (1999)

\bibitem{kosuge2022probabilistic}
Kosuge, H., Xagawa, K.: Probabilistic hash-and-sign with retry in the quantum
  random oracle model. Cryptology ePrint Archive  (2022)

\bibitem{matsumoto1988public}
Matsumoto, T., Imai, H.: Public quadratic polynomial-tuples for efficient
  signature-verification and message-encryption. In: Workshop on the Theory and
  Application of Cryptographic Techniques. pp. 419--453. Springer (1988)

\bibitem{miller1985use}
Miller, V.S.: Use of elliptic curves in cryptography. In: Conference on the
  theory and application of cryptographic techniques. pp. 417--426. Springer
  (1985)

\bibitem{moh1999public}
Moh, T.: A public key system with signature and master key functions.
  Communications in Algebra  \textbf{27}(5),  2207--2222 (1999)

\bibitem{patarin1997oil}
Patarin, J.: The {O}il and {V}inegar signature scheme. In: Dagstuhl Workshop on
  Cryptography September 1997 (1997)

\bibitem{perlner2020rainbow}
Perlner, R., Smith-Tone, D.: Rainbow band separation is better than we thought.
  Cryptology ePrint Archive  (2020)

\bibitem{petzoldt2010cyclicrainbow}
Petzoldt, A., Bulygin, S., Buchmann, J.: {CyclicRainbow}--a multivariate
  signature scheme with a partially cyclic public key. In: International
  Conference on Cryptology in India. pp. 33--48. Springer (2010)

\bibitem{petzoldt2010selecting}
Petzoldt, A., Bulygin, S., Buchmann, J.: Selecting parameters for the {R}ainbow
  signature scheme. In: International Workshop on Post-Quantum Cryptography.
  pp. 218--240. Springer (2010)

\bibitem{Proos_Zalka_2003}
Proos, J., Zalka, C.: Shor's discrete logarithm quantum algorithm for elliptic
  curves. Quantum Inf. Comput.  \textbf{3}(4),  317--344 (2003).
  \doi{10.26421/QIC3.4-3}, \url{https://doi.org/10.26421/QIC3.4-3}

\bibitem{rivest1978method}
Rivest, R.L., Shamir, A., Adleman, L.: A method for obtaining digital
  signatures and public-key cryptosystems. Communications of the ACM
  \textbf{21}(2),  120--126 (1978)

\bibitem{sakumoto2011provable}
Sakumoto, K., Shirai, T., Hiwatari, H.: On provable security of {UOV} and {HFE}
  signature schemes against chosen-message attack. In: International Workshop
  on Post-Quantum Cryptography. pp. 68--82. Springer (2011)

\bibitem{shamir1994efficient}
Shamir, A.: Efficient signature schemes based on birational permutations. In:
  Annual International Cryptology Conference. pp. 1--12. Springer (1994)

\bibitem{shor1994algorithms}
Shor, P.W.: Algorithms for quantum computation: Discrete logarithms and
  factoring. In: Proceedings 35th annual {Symposium on Foundations of Computer
  Science}. pp. 124--134. Ieee (1994)

\bibitem{smith2020rainbow}
Smith-Tone, D., Perlner, R., et~al.: Rainbow band separation is better than we
  thought  (2020)

\bibitem{tao2013simple}
Tao, C., Diene, A., Tang, S., Ding, J.: Simple matrix scheme for encryption.
  In: International Workshop on Post-Quantum Cryptography. pp. 231--242.
  Springer (2013)

\bibitem{thomae2012generalization}
Thomae, E.: A generalization of the rainbow band separation attack and its
  applications to multivariate schemes. Cryptology ePrint Archive  (2012)

\bibitem{wiedemann1986solving}
Wiedemann, D.: Solving sparse linear equations over finite fields. IEEE
  transactions on information theory  \textbf{32}(1),  54--62 (1986)

\bibitem{wolf2006security}
Wolf, C., Braeken, A., Preneel, B.: On the security of stepwise triangular
  systems. Designs, Codes and Cryptography  \textbf{40}(3),  285--302 (2006)

\bibitem{yang2005building}
Yang, B.Y., Chen, J.M.: Building secure tame-like multivariate public-key
  cryptosystems: The new {TTS}. In: Australasian Conference on Information
  Security and Privacy. pp. 518--531. Springer (2005)

\end{thebibliography}
% \begin{thebibliography}{8}
% \bibitem{ref_article1}
% Author, F.: Article title. Journal \textbf{2}(5), 99--110 (2016)

% \bibitem{ref_lncs1}
% Author, F., Author, S.: Title of a proceedings paper. In: Editor,
% F., Editor, S. (eds.) CONFERENCE 2016, LNCS, vol. 9999, pp. 1--13.
% Springer, Heidelberg (2016). \doi{10.10007/1234567890}

% \bibitem{ref_book1}
% Author, F., Author, S., Author, T.: Book title. 2nd edn. Publisher,
% Location (1999)

% \bibitem{ref_proc1}
% Author, A.-B.: Contribution title. In: 9th International Proceedings
% on Proceedings, pp. 1--2. Publisher, Location (2010)

% \bibitem{ref_url1}
% LNCS Homepage, \url{http://www.springer.com/lncs}. Last accessed 4
% Oct 2017
% \end{thebibliography}
\end{document}